\documentclass{svjour3}
\usepackage[utf8x]{inputenc}
\usepackage[american]{babel}  
\usepackage{graphicx}
\usepackage[mathcal]{euscript}
\usepackage{mathrsfs}
\usepackage{hyperref}
\usepackage{amssymb}
\usepackage{bbm}
\usepackage{amsmath}
\usepackage{esint}
\usepackage{natbib} 
\usepackage{caption}
\usepackage{subcaption}
\title{ Exponential Family Mixed Membership Models for Soft~Clustering of Multivariate Data}
\titlerunning{Mixed Membership Models for Soft Clustering}
\date{}
\author{Arthur White \and  Thomas~Brendan~Murphy}
\institute{Trinity~College~Dublin, the University of Dublin, College Green, Dublin 2, Ireland  \and University~College~Dublin, Belfield, Dublin 4, Ireland. \and \email{arwhite@tcd.ie} \and \email{brendan.murphy@ucd.ie}}

\newcommand{\Esub}[2]{{\mathbb E}_{#2}\left[{#1}\right]}

\begin{document}
\maketitle
\begin{abstract}
For several years, model-based clustering methods have successfully tackled many of the challenges presented by data-analysts. However, as the scope of data analysis has evolved, some problems may be beyond the standard mixture model framework. One such problem is when observations in a dataset come from overlapping clusters, whereby different clusters will possess similar parameters for multiple variables. In this setting, mixed membership models, a soft clustering approach whereby observations are not restricted to single cluster membership, have proved to be an effective tool. In this paper, a method for fitting mixed membership models to data generated by a member of an exponential family is outlined. The method is applied to count data obtained from an ultra running competition, and compared with a standard mixture model approach. 

\end{abstract}
\section{Introduction}\label{sec:Intro}
The field of model-based clustering (MBC) \citep{fraley02,mclachlan02} has successfully tackled many of the challenges presented by data-analysts. Within this framework, observations in a dataset are modelled as being drawn from one of several probability distributions. One of the central tenets of MBC, as stated by \citet{fraley02}, is that  datapoints may then be classified so that ``each component probability distribution corresponds to a cluster." While more recent developments, such as those by \citet{baudry10} have evolved this definition somewhat, fundamentally within this framework a clustering solution is sought whereby observations are partitioned into distinct groups, so that observations which have non-negligible posterior probability of belonging to more than one component are seen as having uncertain group membership, and are perhaps indicative of a poorly fitted model.  

However, there are several instances where such a model may prove too restrictive, and it is convenient to introduce a soft clustering approach so that individual observations are modelled by a mixture of components. Examples include: topic modelling, where documents are often interpreted as covering a combination of topics \citep{blei03, erosheva04}; micro cDNA arrays, where overlapping genetic characteristics can be exhibited \citep{Rogers05}; functional disability surveys, where symptoms may be shared \citep{erosheva07} and elections with preferential voting systems, where voters' political positions can viewed as some combination of multiple types \citep{gormley09}\footnote{Note that these examples use different terminology to describe their methods: latent Dirichlet allocation \citep{blei03}, latent process decomposition \citep{Rogers05} and grade of membership  \citep{erosheva07,gormley09}. Each of the models allocate individual observations to multiple components in a similar fashion, which we refer in general to as a mixed membership model \citep{erosheva04}.}. In each of these examples, the cited authors use mixed membership models to analyse the data. Within this framework, observations may be modelled as possessing multiple attributes from the different component probability distributions which are assumed to form the latent structure of the data. Thus, an observation may possess high posterior membership to two or more components with a high degree of certainty. 

The general case of mixed membership models, where quite general component distributions were allowed, has been outlined by \cite{erosheva04}, however, details of how inference is to be performed are omitted; a variational Bayes approximation is recommended, but not described. Other studies \citep{blei03,erosheva04,Rogers05,gormley09} outline a mixed membership approach directly for the problem at hand, and propose to perform inference via either variational Bayes methods \citep{blei03,erosheva07,Rogers05} and/or MCMC schemes~\citep{erosheva07,gormley09}. \citet{Airoldi2006,Airoldi2007} discuss mixed membership models with an emphasis on the issue of model selection. See \citet{Airoldi2014} for a detailed overview of the historical development of mixed membership models and the main areas in which they have been applied. In this paper, the mixed membership approach and a variational Bayes method for inference are outlined for the case where component distributions are members of an exponential family. 

Examples of the method are applied to count data, where the corresponding component distribution is chosen to be Poisson, are provided. The method is first applied to data obtained from a 24 hour ultra running competition, where the hourly number of laps completed by each competitor has been recorded. A comparison is then made to a mixture model approach consistent with standard MBC practices.

The rest of the paper is detailed as follows: Section \ref{sec:MixMem} outlines the general model specification for a mixed membership framework for members of an exponential family. Parameter estimation and model selection, as well as some model evaluation tools and a brief overview of the mixture framework is then discussed in Section \ref{sec:Estimation}. The running data is introduced in Section \ref{sec:24H}, with mixture and mixed membership models fitted to the data and compared. Possible extensions to the model are then discussed in Section \ref{sec:Conclusions}. 

\section{Model Specification}\label{sec:MixMem}
We describe the mixed membership framework. Let $\mathbf{X} = (\mathbf{X}_1, \ldots, \mathbf{X}_N)$ denote our dataset, consisting of $N$ observations of $M$ attributes.  We assume that some number $G$ of \emph{basis profiles} underwrite the data. We use this term to distinguish from terms such as group or cluster, that are commonly used with respect to mixture models. Rather than treating each observation as belonging to a distinct cluster, observations are considered to be some composition of these profiles. 

Weight (or mixed membership) parameters $\boldsymbol{\tau} = (\boldsymbol{\tau}_{1}, \ldots, \boldsymbol{\tau}_{N})$ are assigned to observations $\mathbf{X}$, so that for each $\mathbf{X}_n = (X_{n1}, \ldots, X_{nM})$,  $\boldsymbol{\tau}_n = (\tau_{n1}, \ldots, \tau_{nG})$. Each $\tau_{ng}$ can be interpreted as the probability that an observation will have membership to profile $g$ for an attribute $m$, so each $\tau_{ng} > 0$, $\sum^G_{g=1}  \tau_{ng} =1$. Thus, for a given observation, the \emph{a priori} probability of profile membership is the same for each attribute. Each $\boldsymbol{\tau}_n$ is assumed to follow a Dirichlet distribution, with common hyperparameter $\boldsymbol{\delta} = (\delta_1, \ldots, \delta_G).$

Profile memberships by attribute for $\mathbf{X}$ are denoted by the  the indicator variable~$\mathbf{Z} = (\mathbf{Z}_{1}, \ldots, \mathbf{Z}_{N})$, where $\mathbf{Z}_n = (\mathbf{Z}_{n1}, \ldots, \mathbf{Z}_{nM})$. Specifically, profile membership for each $X_{nm}$ is denoted by the indicator variable~$\mathbf{Z}_{nm} = ({Z}_{nm1}, \ldots, {Z}_{nmG})$, where:

\[ Z_{nmg} = \left\{ \begin{array}{ll}
1 & \mbox{if observation $n$ is member of profile $g$ for attribute $m$;} \\
0 & \mbox{otherwise}.
\end{array} \right. \] 
Each $\mathbf{Z}_{nm}$ is modelled as a multinomial distribution, depending on the probability~$\boldsymbol{\tau}_n$.

Lastly, we use $\boldsymbol{\theta}^{\top} = (\boldsymbol{\theta}_1, \ldots, \boldsymbol{\theta}_M) $, to denote the distribution of data conditional on profile membership, $\boldsymbol{\theta}_m = (\boldsymbol{\theta}_{1m}, \ldots, \boldsymbol{\theta}_{Gm}) $ . For membership to profile $g$ for attribute $m$, $\boldsymbol{\theta}_{gm} $ denotes the underlying parameter(s) of a distribution density $p_1(x_{nm} \mid \boldsymbol{\theta}_{gm})$. We restrict $p_1(x_{nm} \mid \boldsymbol{\theta}_{gm})$ to be a member of an exponential family of distributions: \[p_1(x_{nm} \mid \boldsymbol{\theta}_{gm}) = h(x_{nm}){k}(\boldsymbol{\theta}_{gm})\exp \left \{\mathbf{r}(\boldsymbol{\theta}_{gm}) ^\top\mathbf{s}(x_{nm})\right\},\]
where $\mathbf{r}(\boldsymbol{\theta}_{gm})$ is the natural vector of parameters for $\boldsymbol{\theta}_{gm}$,  $\mathbf{s}(x_{nm})$ the sufficient statistic for $x_{nm}$, and $h(x_{nm})$ is a normalising constant. Note that the dimensions of $\boldsymbol{\theta}_{gm}, \mathbf{s}(x_{nm}), \mbox{ and } \mathbf{r}(\boldsymbol{\theta}_{gm})$ depend on the distribution in question.

The generative process for $\mathbf{X}$ is thus assumed to be given by the following steps: 
\begin{itemize}
\item for each $ n \in 1, \ldots, N: \boldsymbol{\tau}_{n} \sim \mbox{ Dirichlet}(\boldsymbol{\delta})$ \vspace{5mm} 
\item for each $ m \in 1, \ldots, M:  \mathbf{Z}_{nm} \sim \mbox{ Multinomial}(1, \boldsymbol{\tau}_{n})$  \vspace{5mm}
\item $X_{nm} \mid Z_{nmg}=1 \sim p_1(x_{nm} \mid \boldsymbol{\theta}_{gm})$.
\end{itemize}

In the special case where profile distributions are Multinomial(1, $\boldsymbol{\theta}_{gm}$), for all $g, m$, then at an individual level observations will also follow a multinomial distribution, with parameters that are a convex combination of the profile parameters \citep{Galyardt2014}. In the more general case, individuals should be interpreted as switching between profiles across attributes. 


The complete-data posterior for a mixed membership model takes the form:
\begin{equation}
p( \boldsymbol{\tau}, \boldsymbol{\theta}, \mathbf{Z | x}, \boldsymbol{\delta}, \boldsymbol{\eta}, \boldsymbol{\nu}) \propto p_2(\mathbf{x | Z}, \boldsymbol{\theta})p_3(\mathbf{Z} | \boldsymbol{\tau}) p_4(\boldsymbol{\tau | \delta})\prod^G_{g=1}\prod^M_{m=1}p_5(\boldsymbol{\theta} \mid {\eta}_{gm}, \boldsymbol{\nu}_{gm}), \label{eq:ModelPost} \\
\end{equation}
where 
\begin{eqnarray}
p_2(\mathbf{x | Z}, \boldsymbol{\theta}) &=& \prod^{N}_{n=1} \prod^{M}_{m=1} \prod^{ G}_{g=1} p_1(x_{nm} | \boldsymbol{\theta}_{gm})^{Z_{nmg}} \nonumber\\
p_3(\mathbf{Z} | \boldsymbol{\tau}) &=& \prod^{N}_{n=1}\prod^G_{g=1} \tau_{ng}^{\sum_{m=1}^M Z_{nmg}} \nonumber \\
p_4(\boldsymbol{\tau | \delta}) &=& \prod^{N}_{n=1} \frac{\Gamma(\sum^G_{h=1}\delta_h)}{\prod^G_{h=1} \Gamma( \delta_h) } \prod^{G}_{g=1} \tau_{ng}^{\delta_{g} -1} \nonumber \\
p_5(\boldsymbol{\theta}_{gm} \mid {\eta}_{gm}, \boldsymbol{\nu}_{gm}) &=& h({\eta}_{gm}, \boldsymbol{\nu}_{gm}){k}(\boldsymbol{\theta}_{gm})^{{\eta}_{gm}}\exp\{\mathbf{r}(\boldsymbol{\theta}_{gm})^\top \boldsymbol{\nu}_{gm}\}. \nonumber
\end{eqnarray}
where we have assumed conjugate priors for $p_1(\mathbf{x} | \boldsymbol{\theta}) $ and $p_3(\mathbf{Z} | \boldsymbol{\tau})$. 

Note that the form of the posterior outlined in Equation~(\ref{eq:ModelPost}) makes an implicit assumption of the exchangeability of each latent variable $\mathbf{Z}_n$ \citep[see Section 3.1,][]{blei03}. That is, the likelihood of the model will be unchanged for any permutation of the variable index $m=1, \ldots, M.$ Thus, for any observation $\mathbf{X}_n$, all of the observed variables $(X_{n1}, \ldots, X_{nM})$ are assumed to be independent, conditional on their respective profile memberships $(\mathbf{Z}_{n1}, \ldots, \mathbf{Z}_{nM})$. The use of latent variables in a data augmentation approach can also be motivated by a fundamental representation theorem; see \citet[][Section 3]{erosheva07} for further details.

\begin{figure}[t!]
\begin{center}
 \begin{subfigure}[b]{0.49\textwidth}
 \includegraphics[width=0.99\linewidth]{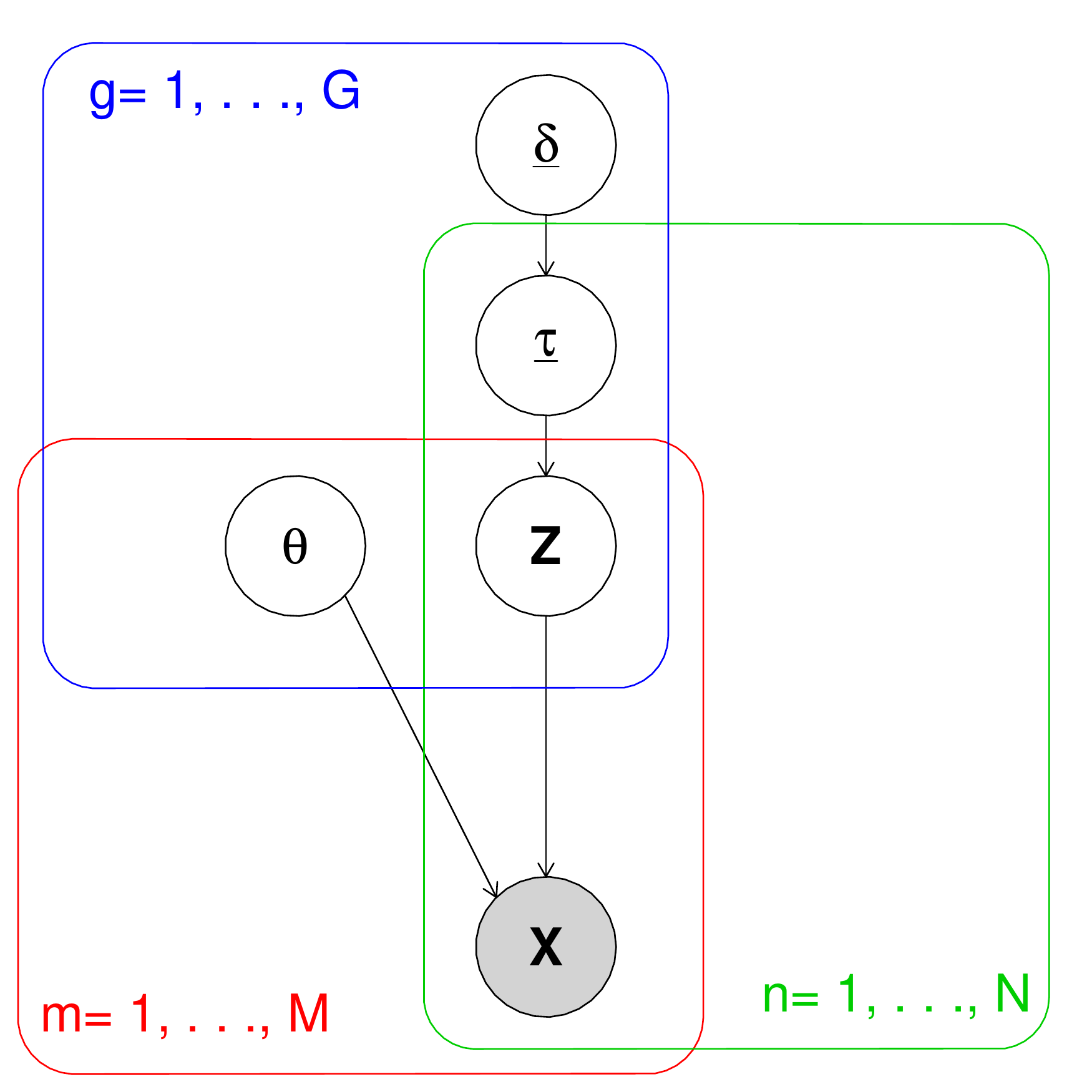}
               \caption{Mixed Membership Model}
                \label{fig:GraphPlota}
         \end{subfigure}
 \begin{subfigure}[b]{0.49\textwidth}
 \includegraphics[width=0.99\linewidth]{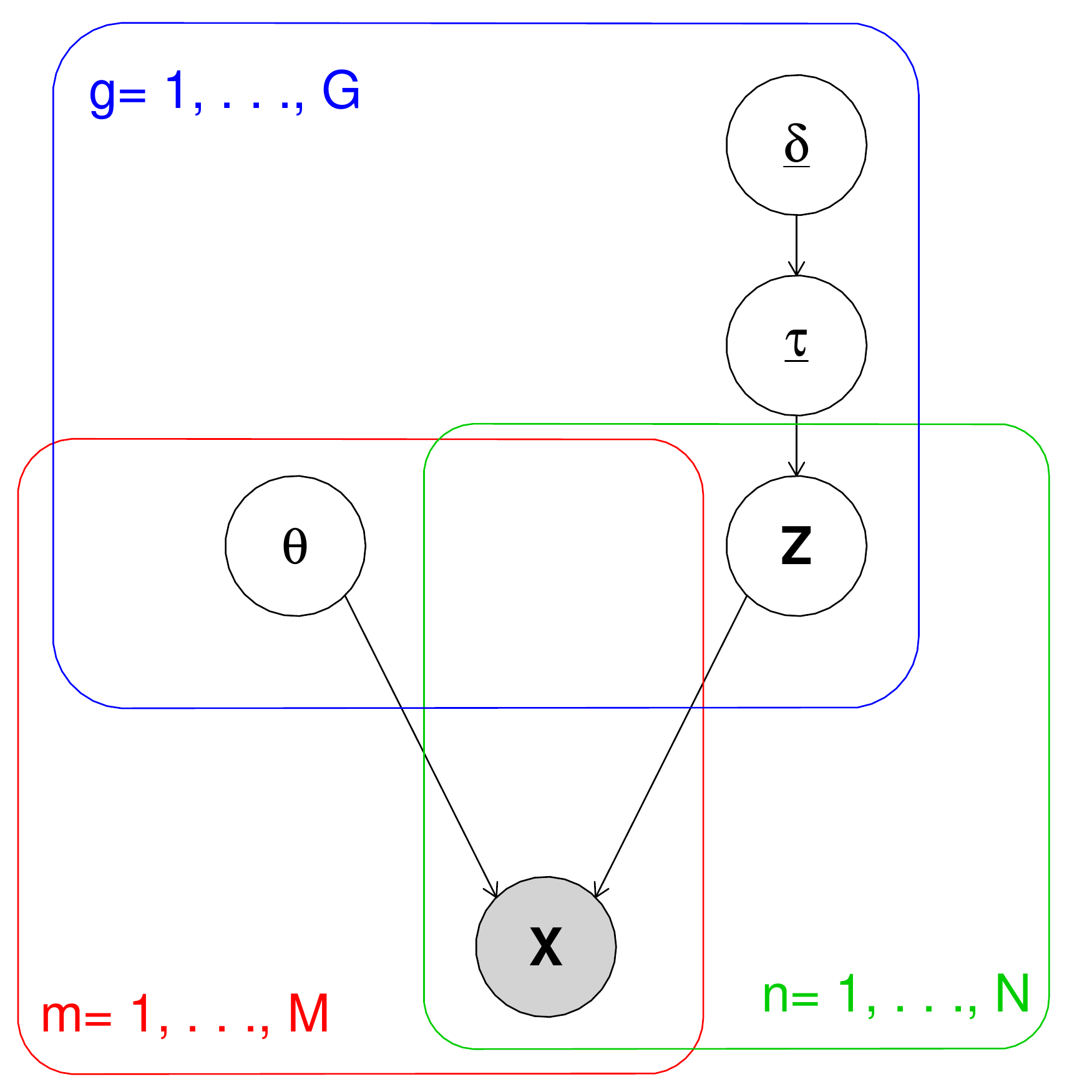}
                \caption{Mixture Model}
                \label{fig:GraphPlotb}
         \end{subfigure}
\caption{Graphical depiction of the mixed membership formulation (a) in comparison to the standard mixture model (b), for the case where $\boldsymbol{\theta}$ is treated as a nuisance parameter. Note in particular the different positions of $\boldsymbol{\tau}$ and $\mathbf{Z}$ with respect to the plate notation.}
\end{center}
\label{fig:GraphPlot}
\end{figure}

A graphical depiction of Equation (\ref{eq:ModelPost}) is shown in Figure \ref{fig:GraphPlota}. For comparison, a mixture model is shown in Figure \ref{fig:GraphPlotb}; this model is formally described in Section~\ref{sec:MixModel}. We repeat notation for the models to highlight similarities in structure. The plate notation in the graph represents the dimensionality of the model parameters. In particular, the different positions of $\boldsymbol{\tau}$ and $\mathbf{Z}$ with respect to this notation illustrate the additional complexity of the mixed membership model.

Note that only the hyperparameter for the prior $p_4(\boldsymbol{\tau | \delta})$ was included in Figure \ref{fig:GraphPlota}, and that the prior $p_5(\boldsymbol{\theta}_{gm} \mid {\eta}_{gm}, \boldsymbol{\nu}_{gm})$ was omitted from the outlined data generative process. This is in keeping with previous studies \citep{blei03,erosheva07,Rogers05} where only $\boldsymbol{\tau}$ has been considered a parameter of interest, with $\boldsymbol{\theta}$ treated as a nuisance parameter, with the prior specification for $\boldsymbol{\eta} \mbox{ and } \boldsymbol{\nu}$ set as small as possible, so that $p_5(\boldsymbol{\theta}_{gm} \mid {\eta}_{gm}, \boldsymbol{\nu}_{gm})$ is as close to a uniform distribution as possible. In either case, calculation of the normalization constant in (\ref{eq:ModelPost}) is intractable \citep{blei03}. For completeness, we consider both cases when discussing the inference method for the model, however, when applying the method to data we choose the nuisance parameter method. While we examine the estimated parameters $\boldsymbol{\hat{\theta}}$ in Section~\ref{sec:24H} in order to interpret the clusters, our primary interest remains the estimation of the underlying mixed membership structure. To perform inference we appeal to variational methods \citep[Chapter 10]{beal03, ormerod10, bishop06}.

\section{Parameter Estimation}\label{sec:Estimation}
In this section parameter estimation for mixed membership exponential family models are outlined. While some of these results are the same as those found in \citep{blei03} the approach as outlined here more closely follows the more general derivation provided in \citet[Chapter 10]{bishop06}.  As a running example, we illustrate how these methods are applied to data generated from a Poisson distribution, i.e., the case where \begin{equation} p_6(x_{nm} \mid\boldsymbol{\theta}_{gm}) = \frac{\exp(-\boldsymbol{\theta}_{gm})\boldsymbol{\theta}_{gm}^{x_{nm}}}{x_{nm}!}. \label{eq:pois} \end{equation} Then $p_6(x_{nm} \mid\boldsymbol{\theta}_{gm})$ is a member of an exponential family with the following specifications:  $h(x_{nm}) =  {1}/{x_{nm}!}, k(\boldsymbol{\theta}_{gm}) = \exp( -\boldsymbol{\theta}_{gm}), s(x_{nm}) = x_{nm}, \mbox{ and } r(\boldsymbol{\theta}_{gm}) = \log(\boldsymbol{\theta}_{gm}).$
A $\mbox{Gamma}(\alpha, \beta)$ distribution is a conjugate prior for a Poisson distribution: \[p_7(\boldsymbol{\theta}_{gm} | \alpha, \beta) = \frac{\beta^{\alpha}}{\Gamma({\alpha})} \boldsymbol{\theta}_{gm}^{\alpha -1} \exp(-\beta\boldsymbol{\theta}_{gm}).\] Matching notation from the previous section gives $\eta =  \beta, \nu = \alpha -1, \mbox{ and } h(\eta, \nu) = {\eta^{\nu +1}}/{\Gamma(\nu + 1)}$. The method applied in Section \ref{sec:24H} also uses this distribution.
\subsection{Variational Bayes}\label{sec:est}

The posterior (\ref{eq:ModelPost}) is approximated using a variational Bayes method \citep{blei03,Rogers05,erosheva07} whereby the posterior $p( \boldsymbol{\tau}, \boldsymbol{\theta}, \mathbf{Z | x}, \boldsymbol{\delta}, \boldsymbol{\eta}, \boldsymbol{\nu})$ is replaced by an approximating set of distributions $q(\mathbf{Z}, \boldsymbol{\tau}, \boldsymbol{\theta} | \boldsymbol{\phi}, \boldsymbol{\gamma}, \boldsymbol{\eta^{\prime}}, \boldsymbol{\nu^{\prime}})$  that factor independently:

\begin{equation}
 q(\mathbf{Z}, \boldsymbol{\tau}, \boldsymbol{\theta} | \boldsymbol{\phi}, \boldsymbol{\gamma}, \boldsymbol{\eta^\prime}, \boldsymbol{\nu^\prime}) =  q_1(\boldsymbol{\tau} | \boldsymbol{\gamma}) q_2(\mathbf{Z} | \boldsymbol{\phi}) q_3(\boldsymbol{\theta} | \boldsymbol{\eta^\prime}, \boldsymbol{\nu^\prime}), \label{eq:LB}
\end{equation}
where $\boldsymbol{\gamma}, \boldsymbol{\phi}, \boldsymbol{\eta^\prime} \mbox{ and } \boldsymbol{\nu^\prime} $ are free variational parameters of $q_1(\boldsymbol{\tau}), q_2(\mathbf{Z}) \mbox{ and } q_3(\boldsymbol{\theta})$ respectively. Note that $\boldsymbol{\phi}, \boldsymbol{\gamma}, \boldsymbol{\eta^\prime} \mbox{ and } \boldsymbol{\nu^\prime} $ have the same dimensionality as $\boldsymbol{\tau}, \mathbf{Z}, \boldsymbol{\eta} \mbox{ and } \boldsymbol{\nu} $ respectively.


To begin with, we obtain an upper bound to the log posterior $\log p_8(\mathbf{x} \mid \boldsymbol{\delta} , \boldsymbol{\eta}, \boldsymbol{\nu})$ in terms of a posterior $q$ with latent parameters $\mathbf{Z}$, $\boldsymbol{\theta}$ and $\boldsymbol{\tau}$.
\begin{eqnarray} 
\log p_8(\mathbf{x} \mid \boldsymbol{\delta} , \boldsymbol{\eta}, \boldsymbol{\nu}) &=& \log \int \int \sum_{\mathbf{Z}} p(\boldsymbol{\tau} , \boldsymbol{\theta}, \mathbf{Z, x} \mid \boldsymbol{\delta} , \boldsymbol{\eta}, \boldsymbol{\nu}) d\boldsymbol{\tau} d\boldsymbol{\theta}\label{eq:logpost}\\ 
&=& \log \int \int \sum_{\mathbf{Z}} \frac{q(\mathbf{Z} ,\boldsymbol{ \tau}, \boldsymbol{\theta}) p(\boldsymbol{\tau} , \boldsymbol{\theta}, \mathbf{Z, x} \mid \boldsymbol{\delta} , \boldsymbol{\eta}, \boldsymbol{\nu}) }{q(\mathbf{Z} ,\boldsymbol{ \tau}, \boldsymbol{\theta})} d\boldsymbol{\tau}  d\boldsymbol{\theta} \nonumber \\ 
&\geq & \int \int \sum_{\mathbf{Z}} q(\mathbf{Z} ,\boldsymbol{ \tau}, \boldsymbol{\theta}) \log p(\boldsymbol{\tau} , \boldsymbol{\theta}, \mathbf{Z, x} \mid \boldsymbol{\delta} , \boldsymbol{\eta}, \boldsymbol{\nu})d\boldsymbol{\tau}d\boldsymbol{\theta} \nonumber\\
& - & \int \int \sum_{\mathbf{Z}}{q(\mathbf{Z} ,\boldsymbol{ \tau}, \boldsymbol{\theta})} \log{q(\mathbf{Z} ,\boldsymbol{ \tau}, \boldsymbol{\theta})} d\boldsymbol{\tau}d\boldsymbol{\theta}, \label{eq:Jen} 
\end{eqnarray}  
where Eq.(\ref{eq:Jen}) is given by Jensen's inequality. It can be shown that the difference between Eq.(\ref{eq:Jen}) and Eq.(\ref{eq:logpost}) is the Kullback-Liebler divergence ${\cal KL} (p(\boldsymbol{\tau} , \boldsymbol{\theta}, \mathbf{Z, x} \mid \boldsymbol{\delta} , \boldsymbol{\eta}, \boldsymbol{\nu}) \| q(\mathbf{Z} ,\boldsymbol{ \tau}, \boldsymbol{\theta}))$. Thus maximising Eq.(\ref{eq:logpost}) amounts to minimising the divergence between the true posterior $p$ and approximate distribution density $q$. 

Introducing the restriction that the approximate distribution density $q(\mathbf{Z}, \boldsymbol{\tau}, \boldsymbol{\theta} | \boldsymbol{\phi}, \boldsymbol{\gamma}, \boldsymbol{\eta^\prime}, \boldsymbol{\nu^\prime}) $ factors independently, it is then possible to maximise Eq. (\ref{eq:Jen}) with respect to $q(\boldsymbol{\tau)}$:
\begin{eqnarray*} 
&~& \int \int \sum_{\mathbf{Z}} q(\mathbf{Z} ,\boldsymbol{ \tau}, \boldsymbol{\theta}) \log p(\boldsymbol{\tau} , \boldsymbol{\theta}, \mathbf{Z, x} \mid \boldsymbol{\delta} , \boldsymbol{\eta}, \boldsymbol{\nu})d\boldsymbol{\tau}d\boldsymbol{\theta} - \int \int \sum_{\mathbf{Z}}{q(\mathbf{Z} ,\boldsymbol{ \tau}, \boldsymbol{\theta})} \log{q(\mathbf{Z} ,\boldsymbol{ \tau}, \boldsymbol{\theta})} d\boldsymbol{\tau}d\boldsymbol{\theta} \\ 
&=&\int q_1(\boldsymbol{\tau}) \int q_3(\boldsymbol{\theta}) \sum_{\mathbf{Z}} q_2(\mathbf{Z} ) \log p(\boldsymbol{\tau} , \boldsymbol{\theta}, \mathbf{Z, x} \mid \boldsymbol{\delta} , \boldsymbol{\eta}, \boldsymbol{\nu})d\boldsymbol{\theta} d\boldsymbol{\tau}  - \int q_1(\boldsymbol{\tau})\log q_1(\boldsymbol{\tau}) d\boldsymbol{\tau}  +\mbox{constant}\\
&=& \int q_1(\boldsymbol{\tau}) \left \{ \Esub{\log p(\boldsymbol{\tau} , \boldsymbol{\theta}, \mathbf{Z, x} \mid \boldsymbol{\delta} , \boldsymbol{\eta}, \boldsymbol{\nu})}{ \mathbf{Z}, \boldsymbol{\theta}} + \mbox{constant} - \log q_1( \boldsymbol{\tau}) \right \}d\boldsymbol{\tau}  + \mbox{constant}\\ 
&=& \int q_1(\boldsymbol{\tau}) \log \left \{ \frac{  \exp \left(  \Esub{\log p(\boldsymbol{\tau} , \boldsymbol{\theta}, \mathbf{Z, x} \mid \boldsymbol{\delta} , \boldsymbol{\eta}, \boldsymbol{\nu})}{ \mathbf{Z}, \boldsymbol{\theta}}+ \mbox{constant} \right) } {q_1( \boldsymbol{\tau})}  \right \}d\boldsymbol{\tau}  + \mbox{constant} \\ 
&=& -{\cal KL} (\exp \left( \Esub{\log p(\boldsymbol{\tau} , \boldsymbol{\theta}, \mathbf{Z, x} \mid \boldsymbol{\delta} , \boldsymbol{\eta}, \boldsymbol{\nu})}{ \mathbf{Z}, \boldsymbol{\theta}} + \mbox{constant} \right) \| q_1(\boldsymbol{\tau})) + \mbox{constant}.
\end{eqnarray*} 

It can thus be shown that maximising Eq. (\ref{eq:Jen}) with respect to $q_1(\boldsymbol{\tau})$ is equivalent to setting   
\begin{eqnarray*} 
q_1(\boldsymbol{\tau}) &\propto&  \exp \left\{ \Esub{\log p(\boldsymbol{\tau} , \boldsymbol{\theta}, \mathbf{Z, x} \mid \boldsymbol{\delta} , \boldsymbol{\eta}, \boldsymbol{\nu})}{ \mathbf{Z}, \boldsymbol{\theta}} \right \}\\ 
&=& \exp \left( \Esub{\log p_2(\mathbf{x \mid Z}, \boldsymbol{\theta})  + \log p_3( \mathbf{Z} \mid \boldsymbol{ \tau})  + \log p_4(\boldsymbol{\tau \mid \delta}) +\log p_5(\boldsymbol{\theta} \mid \boldsymbol{\eta}, \boldsymbol{\nu})  }{ \mathbf{Z}, \boldsymbol{\theta}} \right)\\ 
&\propto & \exp \left( \Esub{\log p_3( \mathbf{Z} \mid  \boldsymbol{\tau}) + \log p_4(\boldsymbol{\tau \mid \delta}) }{ \mathbf{Z}} \right)\\
&=&  \exp \left( \Esub{ \sum_{n=1}^N \sum_{m=1}^M \sum_{g=1}^G Z_{nmg} \log \tau_{ng} + \sum_{n=1}^N \sum_{g=1}^G (\delta_g -1)\log \tau_{ng} }{\mathbf{Z}} \right)\\
&=& \prod_{n=1}^N \prod_{g=1}^G \tau_{ng}^{\sum_{m=1}^M\Esub{  Z_{nmg}}{\mathbf{Z}} + (\delta_g -1)}\\
&=& \prod_{n=1}^N \prod_{g=1}^G \tau_{ng}^{\gamma_{ng} -1},
\end{eqnarray*}  
which we recognise as a Dirichlet distribution, and where we have introduced the variational parameter $\boldsymbol{\gamma}$. 

Similarly, to maximise Eq. (\ref{eq:Jen}) with respect to $q_2(\mathbf{Z})$ set:
\begin{eqnarray*}
q_2(\mathbf{Z})&\propto &  \exp \left( \Esub{\log p(\boldsymbol{\tau} , \boldsymbol{\theta}, \mathbf{Z, x} \mid \boldsymbol{\delta} , \boldsymbol{\eta}, \boldsymbol{\nu}) }{ \boldsymbol{\tau}, \boldsymbol{\theta}}  \right)\\
&\propto & \exp \left( \Esub{\log p_2( \mathbf{X \mid  Z}, \boldsymbol{\theta})}{\boldsymbol{\theta}} +  \Esub{\log p_3(\mathbf{Z} \mid \boldsymbol{\tau} )}{ \boldsymbol{\tau}} \right)\\
&=&  \exp \left(  \sum_{n=1}^N \sum_{m=1}^M \sum_{g=1}^G Z_{nmg}\Esub{\log p_1( x_{nm} | \boldsymbol{\theta}_{gm})}{\boldsymbol{\theta}}+ \sum_{n=1}^N \sum_{m=1}^M \sum_{g=1}^G Z_{nmg} \Esub{\log \tau_{ng} }{\boldsymbol{\tau}} \right)\\
&=& \prod_{n=1}^N \prod_{m=1}^M \prod_{g=1}^G \exp\left \{  \Esub{\log p_1( x_{nm} | \boldsymbol{\theta}_{gm})}{\boldsymbol{\theta}} +\Esub{ \log (\tau_{ng})}{\boldsymbol{\tau}} \right \}^{ Z_{nmg}}\\
&=& \prod_{n=1}^N \prod_{m=1}^M \prod_{g=1}^G \phi_{nmg}^{ Z_{nmg}}.
\end{eqnarray*}  
This can be recognised as a multinomial distribution, with the variational parameter $\boldsymbol{\phi}$. 

%
The variational approximation $q_3(\boldsymbol{\theta}_{gm})$ has the form:
\begin{eqnarray*}
q_3(\boldsymbol{\theta}_{gm}) &\propto & \exp\left \{\Esub{ \sum^N_{n=1} \log p_2(x_{nm} | \boldsymbol{\theta}_{gm}, Z_{nmg}) }{Z} + \log p_5(\boldsymbol{\theta}_{gm} | \eta_{gm},\boldsymbol{ \nu}_{gm})  \right \} \\
&=& \exp\left \{\sum^N_{n=1} \Esub{  Z_{nmg} \log p_1(x_{nm} | \boldsymbol{\theta}_{gm})}{Z} + \log p_5(\boldsymbol{\theta}_{gm} | \eta_{gm}, \boldsymbol{\nu}_{gm}  \right \} \\
&=& \exp\left \{\sum^N_{n=1}   \Esub{  Z_{nmg}}{\mathbf{Z}} \left ( \log k(\boldsymbol{\theta}_{gm}) +  \log h(x_{nm}) + \mathbf{r}(\boldsymbol{\theta}_{gm})^{\top} \mathbf{s}(x_{nm}) \right ) \right . \\
& + & \left . \log h(\eta_{gm}, \boldsymbol{\nu}_{gm}) + \eta_{gm} \log k(\boldsymbol{\theta}_{gm}) + \mathbf{r}(\boldsymbol{\theta}_{gm})^{\top}\boldsymbol{\nu}_{gm} \right \} \\
&\propto& \exp\left \{\left(\sum^N_{n=1} \Esub{  Z_{nmg}}{\mathbf{Z}}  + \eta_{gm} \right) \log k(\boldsymbol{\theta}_{gm}) + \mathbf{r}(\boldsymbol{\theta}_{gm})^{\top} \left( \sum^N_{n=1} \Esub{  Z_{nmg}}{\mathbf{Z}}  \mathbf{s}(x_{nm}) + \boldsymbol{\nu}_{gm} \right ) \right \} \\
&=& k(\boldsymbol{\theta}_{gm})^{\sum^N_{n=1} \Esub{  Z_{nmg}}{\mathbf{Z}}  + \eta_{gm}} \exp\left \{ \mathbf{r}(\boldsymbol{\theta}_{gm})^{\top} \left( \sum^N_{n=1} \Esub{  Z_{nmg}}{\mathbf{Z}}  \mathbf{s}(x_{nm}) + \boldsymbol{\nu}_{gm} \right ) \right \} \\
&=&k(\boldsymbol{\theta}_{gm})^{\eta^{\prime}_{gm}} \exp\left \{ \mathbf{r}(\boldsymbol{\theta}_{gm})^{\top} \boldsymbol{\nu^{\prime}}_{gm}  \right \} \\
&=& p_5(\boldsymbol{\theta}_{gm} | \eta^{\prime}_{gm}, \boldsymbol{\nu^\prime}_{gm}),
\end{eqnarray*}
where we have introduced the variational parameters ${\eta^\prime} $ and $\boldsymbol{\nu^\prime}$. Thus $q_3(\boldsymbol{\theta}_{gm}) $ will be the a member of the same exponential family as the prior $p_5(\boldsymbol{\theta}_{gm} | {\eta}_{gm}, \boldsymbol{\nu}_{gm})$.

Parameter updates in terms of these variational parameters are as follows:
\begin{eqnarray*}
\gamma_{ng} 
&=&  \sum_{m=1}^M \phi_{nmg} + \delta_g;\\
\phi_{nmg} 
&=& \exp \left \{ \Esub{ \log k(\boldsymbol{\theta}_{gm}) +  \log h(x_{nm}) + \mathbf{r}(\boldsymbol{\theta}_{gm})^{\top} \mathbf{s}(x_{nm}) }{\boldsymbol{\theta}}  + \Psi( \gamma_{ng}) - \Psi\left(\sum^G_{h=1} \gamma_{nh} \right)  \right \} \\
&\propto & \exp \left \{ \Esub{ \log k(\boldsymbol{\theta}_{gm}) + \mathbf{r}(\boldsymbol{\theta}_{gm})^{\top} \mathbf{s}(x_{nm}) }{\boldsymbol{\theta}}  + \Psi( \gamma_{ng})  \right \}; \\
{\eta^\prime}_{gm}
&=& \sum^N_{n=1} \phi_{nmg}  + {\eta}_{gm}; \\
\boldsymbol{\nu^\prime}_{gm} 
&=& \sum^N_{n=1} \phi_{nmg}\mathbf{s}(x_{nm}) + \boldsymbol{\nu}_{gm},
\end{eqnarray*}
where $\Psi$ denotes the digamma distribution \citep{abramowitz}.

In the case of Poisson/Gamma distributed data, the updates for $\boldsymbol{\phi}, \boldsymbol{\nu^\prime}\mbox{ and }\boldsymbol{\eta^\prime} $ become: 
\begin{eqnarray}
\phi_{nmg} &\propto& \exp \left ( \frac{\nu^\prime_{gm} +1}{\eta^\prime_{gm}} + (\Psi(\nu^\prime_{gm} + 1) - \log(\eta^\prime_{gm}))X_{nm} + \Psi(\gamma_{ng}) \right ); \nonumber\\
\eta^\prime_{gm} &=& \sum^N_{n=1} \phi_{nmg} + \beta; \nonumber\\
\nu^\prime_{gm} &=& \sum^N_{n=1} \phi_{nmg}x_{nm} + \alpha -1. \nonumber
\end{eqnarray}

\subsubsection*{Nuisance Parameter}
When treated as a nuisance parameter, the parameter update for $\boldsymbol{\theta}$ can be obtained by direct maximum likelihood estimation of Equation (\ref{eq:ModelPost}). In this case, the log posterior becomes 
\begin{equation*} 
\log p_9(\mathbf{x} \mid \boldsymbol{\delta} , \boldsymbol{\theta}) \geq  \int \sum_{\mathbf{Z}} q_{4}(\mathbf{Z} ,\boldsymbol{ \tau}) \log p_9(\mathbf{x} \mid \boldsymbol{\delta} , \boldsymbol{\theta})d\boldsymbol{\tau} - \int \sum_{\mathbf{Z}}{q_4(\mathbf{Z} ,\boldsymbol{ \tau})} \log{q_4(\mathbf{Z} ,\boldsymbol{ \tau})} d\boldsymbol{\tau}.
\end{equation*}  
The form of $q_1(\boldsymbol{\tau})$ and update for $\boldsymbol{\gamma}$ remain unchanged. While the form of $q_2(\mathbf{Z})$ is the same, the calculation of $\mathbf{\phi}$ differs, however:
\begin{eqnarray*}
q_2(\mathbf{Z})
&\propto & \exp \left( \log p_2( \mathbf{x \mid  Z}, \boldsymbol{\theta}) +  \Esub{\log p_3(\mathbf{Z} \mid \boldsymbol{\tau} )}{ \boldsymbol{\tau}} \right)\\
&=& p_2( \mathbf{X \mid  Z}, \boldsymbol{\theta}) \times  \exp \left( \Esub{\log p_3(\mathbf{Z} \mid \boldsymbol{\tau} )}{ \boldsymbol{\tau}} \right) \\
&=&  \prod_{n=1}^N \prod_{m=1}^M \prod_{g=1}^G \left \{ p_1( x_{nm} \mid \boldsymbol{\theta}_{gm}) \times  \exp \left( \Esub{\log {\tau}_{ng}}{ \boldsymbol{\tau}} \right)  \right \}^{Z_{nmg}}.\\
\end{eqnarray*}
Thus the update for $\boldsymbol{\phi}$ becomes $\phi_{nmg} \propto p_1( x_{nm} \mid \boldsymbol{\theta}_{gm}) \times  \exp \left\{ \Psi({\tau}_{ng}) \right\}.$

The maximum likelihood estimate $\boldsymbol{\hat{\theta}}$ is achieved by solving
\begin{equation*}
 \left . \sum^N_{n=1} \nabla\log p_3({x}_{nm} \mid \boldsymbol{{\theta}}_{gm}, {Z}_{nmg}) \right |_{\boldsymbol{\theta}_{gm} = \boldsymbol{\hat{\theta}}_{gm}}  = 0.\\
\end{equation*}
Substituting in the estimate $\boldsymbol{\phi}$ for $\mathbf{Z}$, and noting that \begin{eqnarray*}  \sum^N_{n=1} \nabla\log p_3(x_{nm} \mid \boldsymbol{\theta}_{gm}, Z_{nmg}) &=&   \sum^N_{n=1} \phi_{nmg} \nabla\log p_1(x_{nm} \mid \boldsymbol{{\theta}}_{gm})  \\
&=& \sum^N_{n=1} \phi_{nmg} \left ( \frac{\nabla k(\boldsymbol{\theta}_{gm}) }{k(\boldsymbol{\theta}_{gm})} + \nabla \mathbf{r}(\boldsymbol{\theta}_{gm})^{\top} \mathbf{s}(x_{nm})  \right ),
\end{eqnarray*}
an estimate of $\boldsymbol{\hat{\theta}}$ can then be obtained by solving: $$-\nabla \mathbf{r}^{-1}(\boldsymbol{\hat{\theta}}_{gm}) \frac{\nabla k(\boldsymbol{\hat{\theta}}_{gm}) }{k(\boldsymbol{\hat{\theta}}_{gm})} =  \frac{\sum^N_{n=1} \phi_{nmg}\mathbf{s}(x_{nm})}{ \sum^N_{n=1} \phi_{nmg}}.$$


In the case of the Poisson distribution this becomes:
\begin{eqnarray} \phi_{nmg} &\propto& \left . \exp(-{\theta}_{gm}){{\theta}_{gm}^{x_{nm}}} \right . \times \exp \left (\Psi(\gamma_{ng}) \right ) \nonumber\\
\hat{\theta}_{gm} &=& \frac{\sum^N_{n=1}\phi_{nmg}x_{nm}}{\sum^N_{n=1}\phi_{nmg}}. \nonumber
\end{eqnarray}

In addition to estimating the profile memberships $\mathbf{Z}$ and model parameters $\boldsymbol{\tau} \mbox{ and } \boldsymbol{\theta}$, \citet{erosheva07,Airoldi2006,Airoldi2007} propose to estimate the hyperparameter $\boldsymbol{\delta}$ using an empirical Bayes method. We omit this step from our analysis, and in the data analysis described in Section~\ref{sec:24H}, we set $\delta_g = 1/G$, for all~$g.$

\subsection{Model Selection and Likelihood Estimation}
While model assumptions require the number of profiles $G$ to be fixed and known, in reality this is not the case. We therefore run the model over a range of values of $G^\prime = 1, \ldots, G^{\max}$, and compare the models post-hoc. While \citet{Airoldi2006} use the variational approximation to Equation (\ref{eq:logpost}) as a surrogate for the Bayesian Information Criterion (BIC) \citep{schwarz1978}, in our opinion, the fact that the approximation (\ref{eq:LB}) provides only a lower bound to the model posterior (\ref{eq:ModelPost}) makes the use of such a criterion difficult to interpret. 

 \citet{Rogers05} propose evaluating the hold-out likelihood of the model, which involves integrating $\boldsymbol{\tau}$ and $\mathbf{Z}$ from the complete-data posterior given in (\ref{eq:ModelPost}). In the case of the Poisson distribution with $\boldsymbol{\theta}$ a nuisance parameter, this becomes:
\begin{equation} p_9(\mathbf{x} | \boldsymbol{\theta}, \boldsymbol{\delta}, G^{\prime}) \propto \prod^N_{n=1} \left \{ \int_{\boldsymbol{\tau_n}} \prod_{m=1}^M  \sum_{g=1}^{G^{\prime}}  \tau_{ng}  \frac{\exp\{-\boldsymbol{\theta}_{gm}\} \boldsymbol{\theta}_{gm}^{x_{nm}}}{x_{nm}!} p(\tau_{ng} | \delta_g) d\boldsymbol{\tau}_n  \right \}\label{eq:MMeq}.
\end{equation}  
Equation (\ref{eq:MMeq}) may be approximated using a Monte Carlo method, by averaging over $T$ draws from the prior $p(\boldsymbol{\tau} | \boldsymbol{\delta})$:
\begin{equation*}
p_9(\mathbf{x} | \boldsymbol{\theta}, \boldsymbol{\delta}, G^{\prime}) \approx \prod^N_{n=1} \left \{ \frac{1}{T} \sum^T_{t=1} \prod_{m=1}^M  \sum_{g=1}^{G^{\prime}}  \tau^{(t)}_{g}  \frac{\exp\{-\boldsymbol{\theta}_{gm}\} \boldsymbol{\theta}_{gm}^{x_{nm}}}{x_{nm}!} \right \}.
\end{equation*}

\subsection{Model Evaluation}\label{sec:SumStat}
While parameter estimates are used to interpret the model fitted in Section \ref{sec:24H}, we also make use of the following statistics, which further help to summarise the data. For convenience these are briefly described here.  

\begin{description}
\item[{\bf Extent of profile membership (EoM)}] The extent to which an observation's attributes appear to be generated by multiple profiles can be estimated using a measure such as EoM \citep{hill73,white12}, where $\mbox{EoM}_n  = \exp (H(\boldsymbol{\hat{\tau}}_n)),$ and $H$ denotes the entropy function, $H(\boldsymbol{\hat{\tau}}_n) = -\sum^G_{g=1}\hat{\tau}_{ng}\log\hat{\tau}_{ng}.$ This estimates the number of profiles from which an observation's variables seem to be drawn. Thus considering the EoM over all observations gives an idea of the amount of mixed membership taking place in the data.


\item[{\bf Maximum \emph{a posteriori} $(\mathbf{\hat{Z}})$} ] We can impose a hard clustering by mapping individuals to their most probable profile memberships for each attribute by setting  ${\hat{Z}}_{nm} = \arg\max_{g = 1, \ldots, G} \left \lbrace {\mathbb P}(\mbox{profile } g | x_{nm})\right \rbrace,$ where $ \mathbb P(\mbox{ profile } g| x_{nm})$, the probability that the observed value $x_{nm}$ results from profile $g$, is estimated by $\hat{\phi}_{nmg}.$ 

It can be shown that every mixed membership model can be re-expressed as a finite mixture model with a much larger number of components~\citep{erosheva07,Galyardt2014}. In effect, these components consist of the distinct permutations of profile membership which occur across attributes in the data. One can think of the profile mapping summary statistic $\mathbf{\hat{Z}}$ as an estimate of this quantity. 

We use the notation $\{ a, b\}$ to indicate the set of individuals whose assigned membership across attributes is some (repeated) permutation of profiles $a$ and $b$. In other words, an observation $n$ is an element of $\{a,b\}$, if $a$ and $b$ are the unique elements in $\mathbf{\hat{Z}}_{n}$.  Note that this notation can be used for any number of profiles: for example, $\{ 1 \}$ indicates the individuals who exclusively map to profile 1 across all attributes.

\item[{\bf Classification uncertainty (}$\mathbf{U}${\bf)}] Another way to scrutinise classification is to consider the uncertainty associated with an observation's profile assignment for each of their attributes \citep{bensmail97}:
$U_{nm} = \min_{g = 1, \ldots, G} \lbrace 1 - {\mathbb P}(\mbox{profile } g | x_{nm}) \rbrace,$
where the lower the uncertainty, the better the classification.
\end{description}

\subsection{Mixture Model Framework}\label{sec:MixModel} 
In Section \ref{sec:24H} the mixed membership approach is compared to the standard MBC approach. To fit a model using the mixture model framework \citep{everitt1981}, we first assume a fixed number $G$ of groups underly the data. We use this term exclusively for mixture models. Let $\boldsymbol{\tau}^{\mbox{mix}} = ({\tau}^{\mbox{mix}}_1, \ldots, {\tau}^{\mbox{mix}}_G)$ denote the prior probability that an observation belongs to each group. Consequently, the likelihood $p_{\mbox{mix}}(\mathbf{x} \mid \boldsymbol{\theta}^{\mbox{mix}}, \boldsymbol{\tau}^{\mbox{mix}})$ then takes the form
$$
p_{\mbox{mix}}(\mathbf{x} \mid \boldsymbol{\theta}^{\mbox{mix}}) = \prod_{n=1}^N \sum_{g=1}^G {\tau}^{\mbox{mix}}_g \prod_{m=1}^M  p_6(x_{nm} \mid {\theta}_{gm}^{\mbox{mix}}),
$$
where $p_6(x_{nm} \mid {\theta}_{gm}^{\mbox{mix}})$ is defined as in Equation \ref{eq:pois}. Direct inference of this likelihood is difficult, but can be facilitated with the introduction of missing data $\mathbf{Z}^{\mbox{mix}} = (\mathbf{Z}^{\mbox{mix}}_1, \ldots, \mathbf{Z}^{\mbox{mix}}_N)$, and $\mathbf{Z}^{\mbox{mix}}_{n} = (Z^{\mbox{mix}}_{n1}, \ldots, Z^{\mbox{mix}}_{nG}),$ for each $n = 1, \ldots, N.$ We define $$ Z^{\mbox{mix}}_{ng} = \left\{ \begin{array}{ll}
1 & \mbox{if observation $n$ is member of Group $g$;} \\
0 & \mbox{otherwise}.
\end{array} \right. $$
From a clustering perspective, each $ Z^{\mbox{mix}}_{n}$ can be interpreted as a latent variable indicating cluster membership \citep{fraley02}. Note that within the mixture model framework, conditional on group membership, observations are assumed to be drawn independently.

We can use similar summary statistics to evaluate the clustering performance of a mixture model to those described in Section~\ref{sec:SumStat}. In particular, define 
${\hat{Z}^{\mbox{mix}}}_{n} = \arg\max_{g = 1, \ldots, G} \left \lbrace {\mathbb P}(\mbox{group } g | \mathbf{x}_{n})\right \rbrace,$ and 
$U^{\mbox{mix}}_{n} = \min_{g = 1, \ldots, G} \lbrace 1 - {\mathbb P}(\mbox{group } g | \mathbf{x}_{n}) \rbrace.$ These map individual observations to groups and assess the uncertainty of this classification respectively. Note that these values assign a single value to each observation (across all attributes), as opposed to the statistics for mixed membership, which potentially assign different values to an observation's attributes.

We omit further details of how inference is performed, except to mention that parameter estimates may be obtained using an EM algorithm \citep{dempster77}. To determine the optimal number of clusters in the data, the model was run over a large number of groups, and the BIC was used to identify the optimal number to fit to the data. While the regularity conditions required for the BIC are not met when choosing the number of groups for a mixture model \citep{biernacki2000}, at a practical level it has proved useful on many occasions \citep{fraley02}. To perform inference in a Bayesian setting, conjugate priors can be chosen in a similar fashion to those already described. The use of priors with different (sensible) choices of hyper-parameters were found to have little effect on the clustering obtained by the application in Section \ref{sec:24H}.

\section{International Association of Ultrarunners 24 Hour World Championships}\label{sec:24H}
\begin{figure}[t!]
\begin{center}
\begin{subfigure}[b]{0.49\textwidth}
 \includegraphics[width=0.99\textwidth]{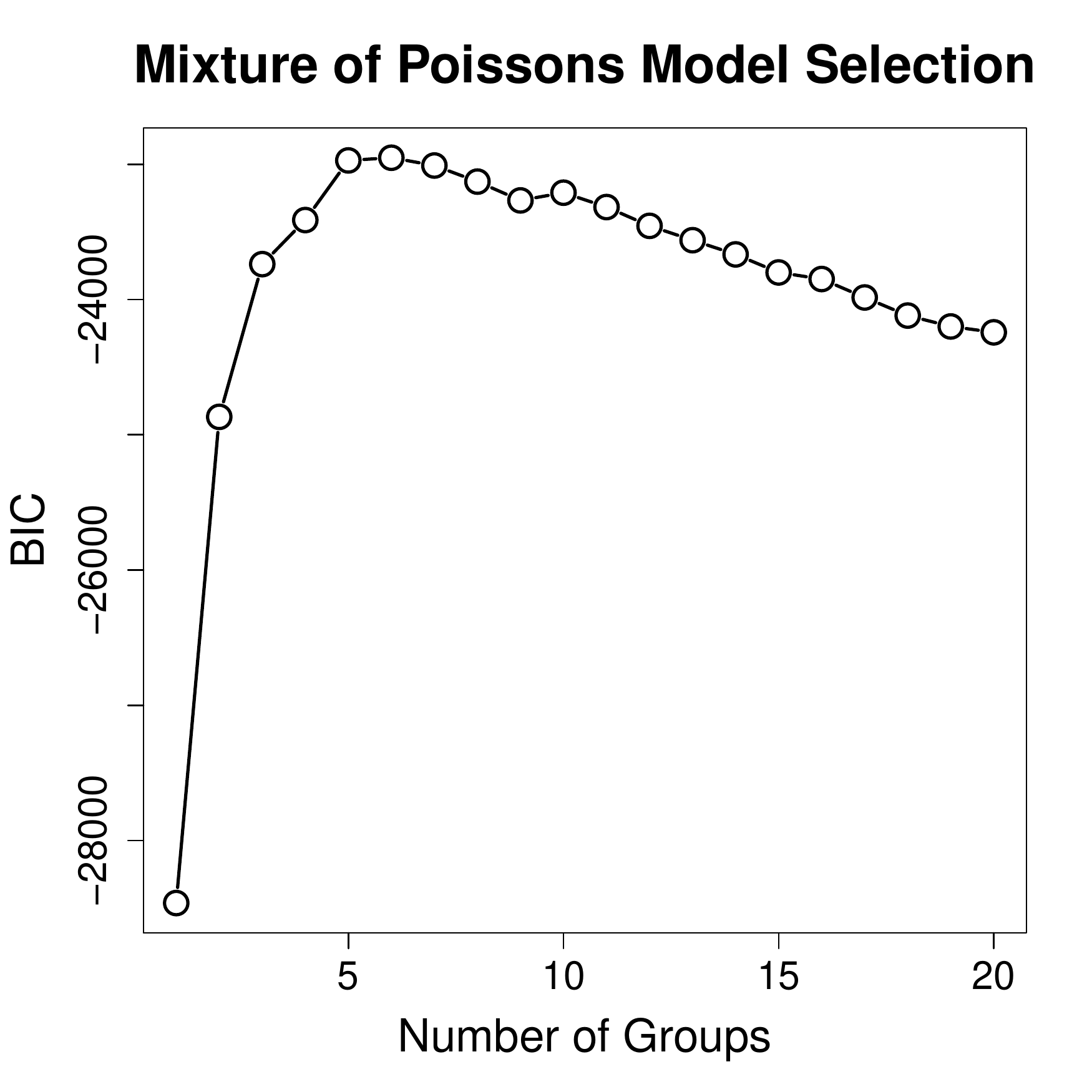}
                \caption{~}
                \label{fig:24Hselecta}
   \end{subfigure}
   \begin{subfigure}[b]{0.49\textwidth}
\includegraphics[width=0.99\textwidth]{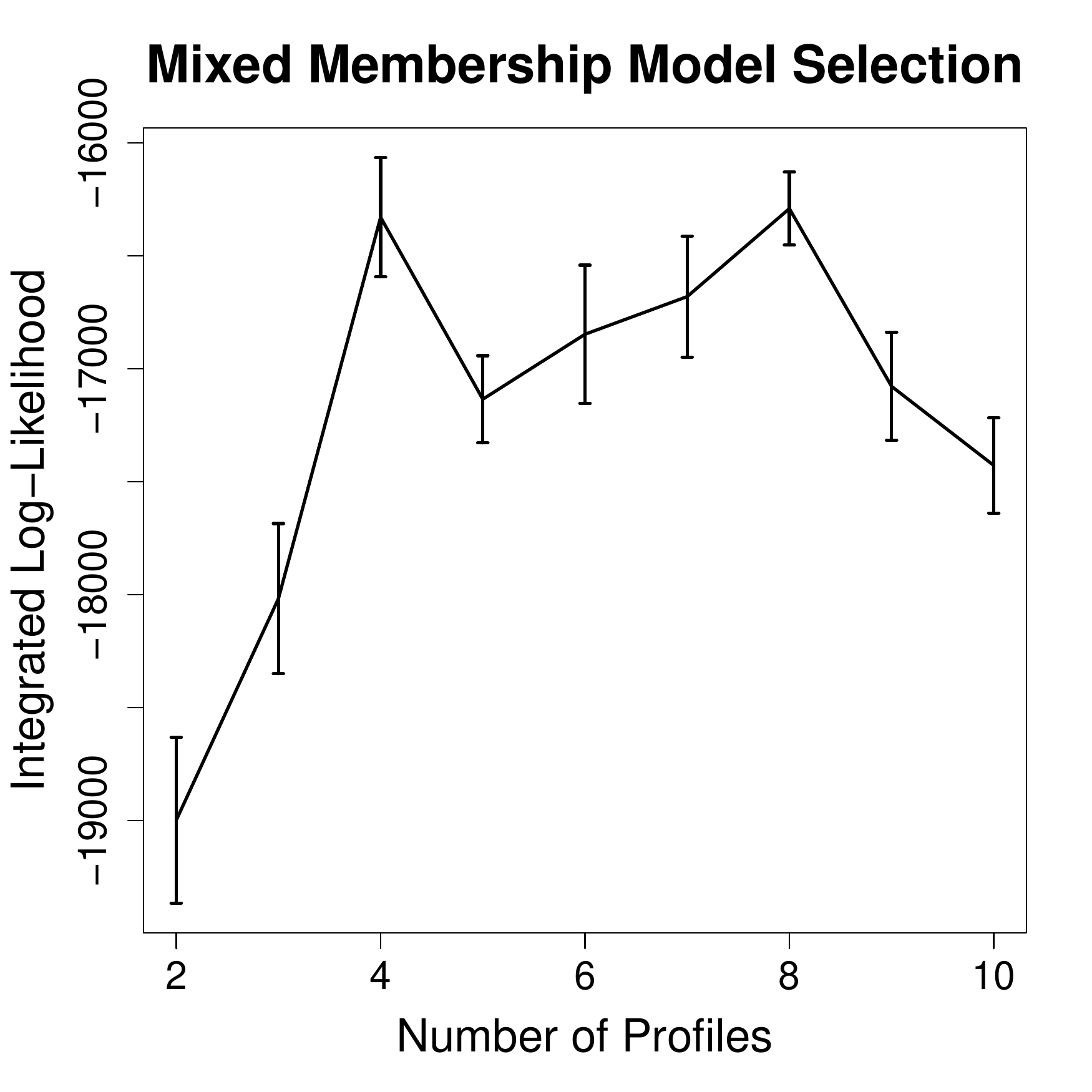}
                \caption{~}
                \label{fig:24Hselectb}
  \end{subfigure}
\end{center}
\caption{BIC (a) and hold-out likelihood (b) values for mixture and mixed membership models applied to the IAU running data. Within the respective frameworks, 6 component (group) and 4 profile models appear to fit the data optimally.
}
\label{fig:24Hselect}
\end{figure}

The International Association of Ultrarunners (IAU) 24 hour World Championships were held in Katowice, Poland on September 8th to 9th, 2013. Two hundred and sixty athletes representing twenty four countries entered the race, which was held on a course consisting of a 1.554~km looped route. An update of the number of laps covered by each athlete was recorded approximately every hour\footnote{A version of this data is available at \url{http://mathsci.ucd.ie/\textasciitilde
brendan/data/24H.xlsx}}. 



Note that the sequential nature of the data means that the exchangeability assumption required by the mixed membership model discussed in Section \ref{sec:MixMem}, as well as the conditional independence assumption required by the mixture model, may both be somewhat unrealistic in this setting. Nevertheless, the approaches appear to identify interesting behaviour in the data, and serve to illustrate important differences between the methods. Both mixture and mixed membership models were applied to the dataset, with the BIC and hold out likelihood suggesting that 6-component and 4-profile fits were optimal; this is illustrated in Figure \ref{fig:24Hselect}.


\subsection{Mixture Model Application}
\begin{figure}[th]
\begin{center}
\begin{subfigure}[b]{0.49\textwidth}
\includegraphics[width=0.99\textwidth]{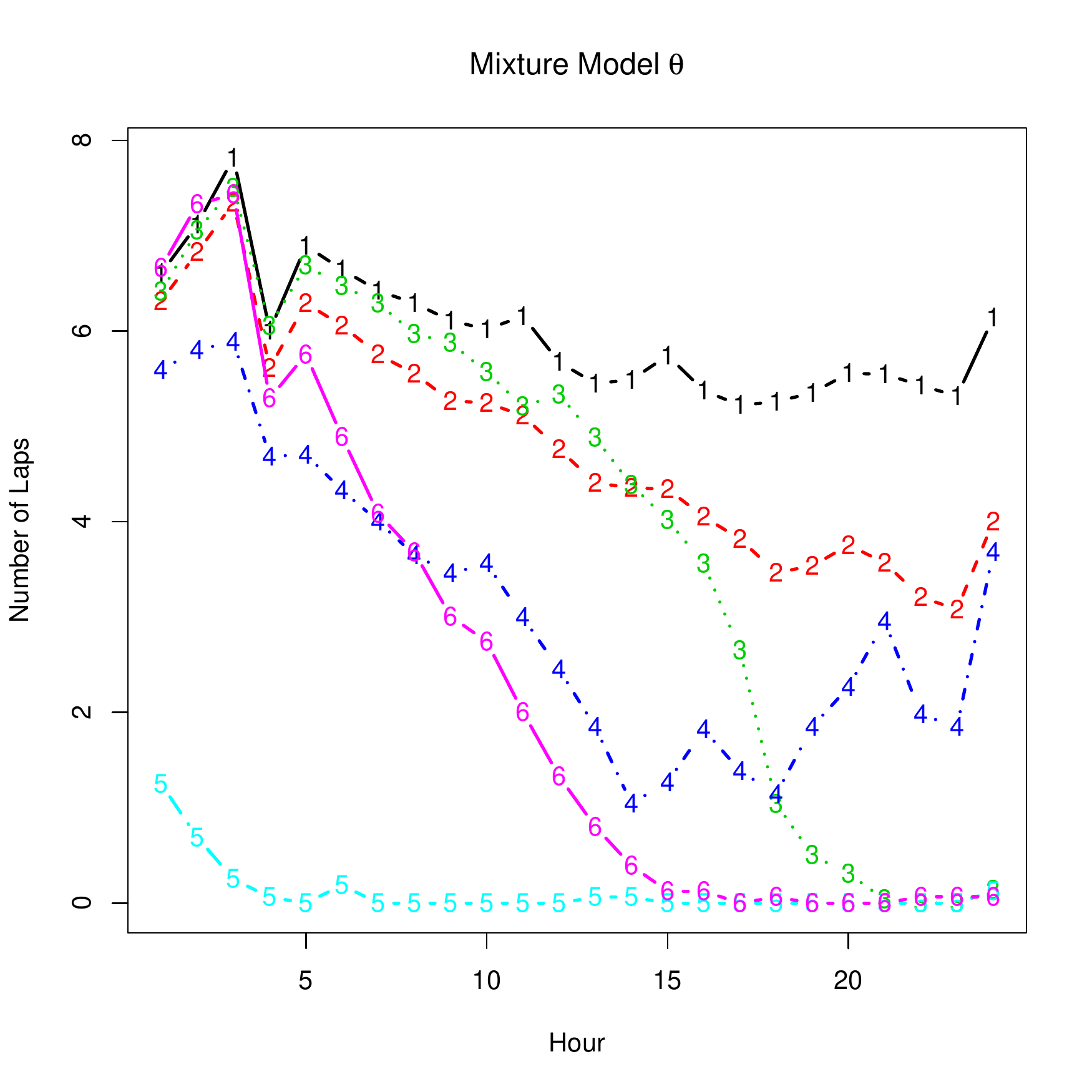}
                \caption{~}
                \label{fig:24Hthetaa}
 \end{subfigure}
\begin{subfigure}[b]{0.49\textwidth}
 \includegraphics[width=0.99\textwidth]{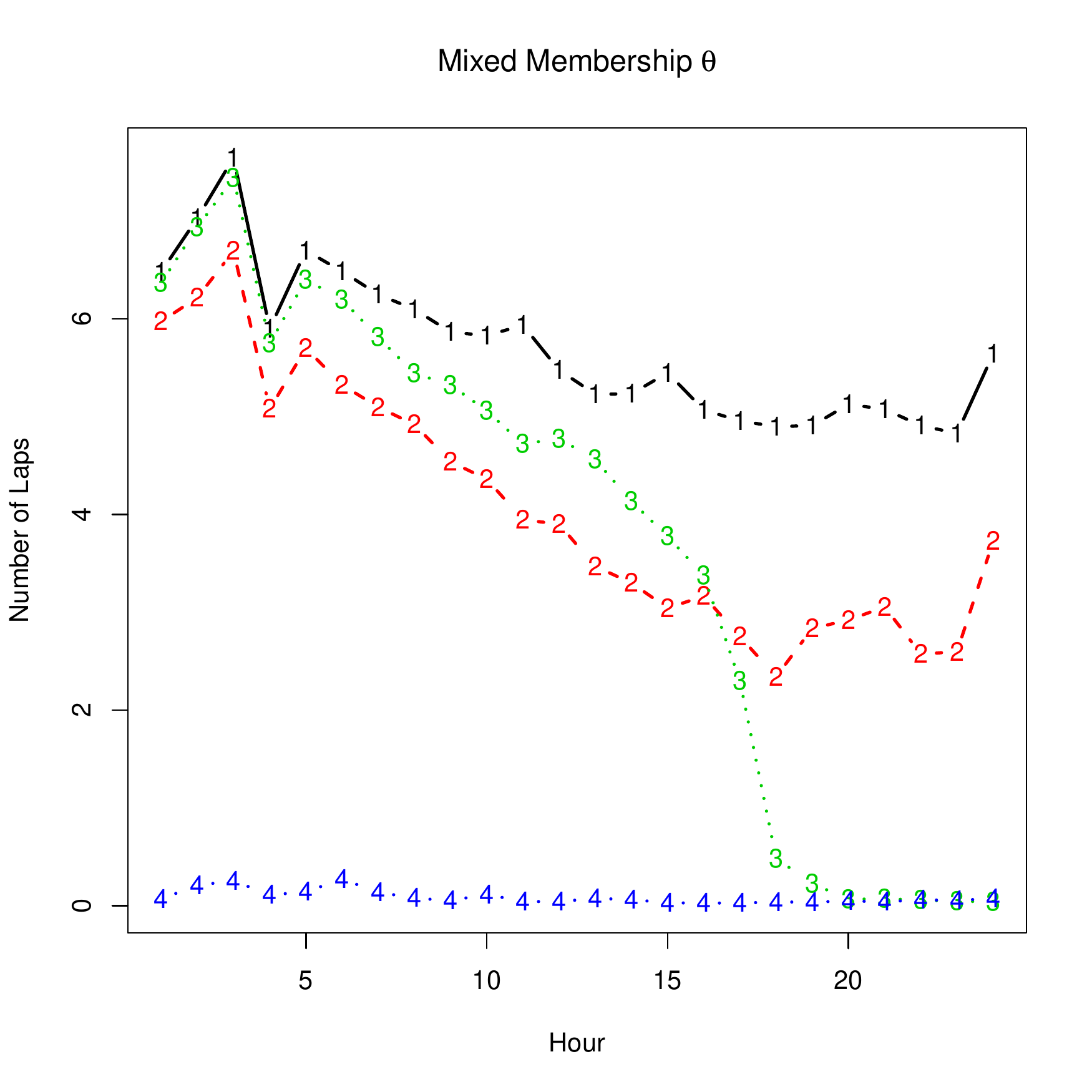}
                \caption{~}
                \label{fig:24Hthetab}
 \end{subfigure}
\end{center}
\caption{Plots of the expected number of  laps completed per hour, conditional on group  and profile membership, for (a) mixture model with 6 components (groups) and (b) mixed membership model with 4 profiles respectively.  }
\label{fig:24Htheta}
\end{figure}

The estimated weight parameters for the 6-component mixture model were \\$\boldsymbol{\tau}^{\mbox{mix}} = (0.40 ~0.33 ~0.08 ~0.07 ~0.06~ 0.06)$. The estimated values  of  $\boldsymbol{\theta}^{\mbox{mix}}$ are  illustrated in Figure \ref{fig:24Hthetaa}. This figure suggests that the two largest groups (Groups 1 and 2) in the dataset ran at a reasonably steady rate over the course of the race, with  Group 2's pace declining in a slightly more pronounced manner during the second half of the race. Three of the four remaining smaller groups, Groups 3,4, and 6, began the race at a similarly high pace to Groups 1 and 2, but were unable to sustain such a rate over the duration of the race. In particular, runners in Groups 3 and 6 failed to complete many laps beyond the 18 and 12 hour marks respectively, while runners clustered in Group 4 maintained a steadier pace throughout the race, and actually improved slightly over the final four hours. Finally, Group 5 consisted of entrants who completed only a very small number of laps over the course of the race, including several runners who completed no laps; this includes race entrants who failed to participate on the day of the race. 

\subsection{Mixed Membership Model Application}

The estimated values  of  $\boldsymbol{\theta}$ for the 4-profile mixed membership model are  illustrated in Figure \ref{fig:24Hthetab}. Based on this plot, profile behaviour conveys much of the same information as the mixture model: over the course of the race, the characteristic behaviour of Profile 1 is to perform at a  high and steady rate; Profile 2 is at a similarly steady but slower pace; Profile 3 begins brightly but declines sharply by the final quarter of the race; while Profile 4 can be characterised as exhibiting extremely low-level, non-participatory behaviour. For convenience, we refer to Profiles 1 to 4 by the following names: Fast Pace, Slow Pace, Rapid Decline, and Non-Participation, respectively. 

\begin{figure}[!t]
\begin{center}
\begin{subfigure}[b]{0.49\textwidth}
\includegraphics[width=0.99\textwidth]{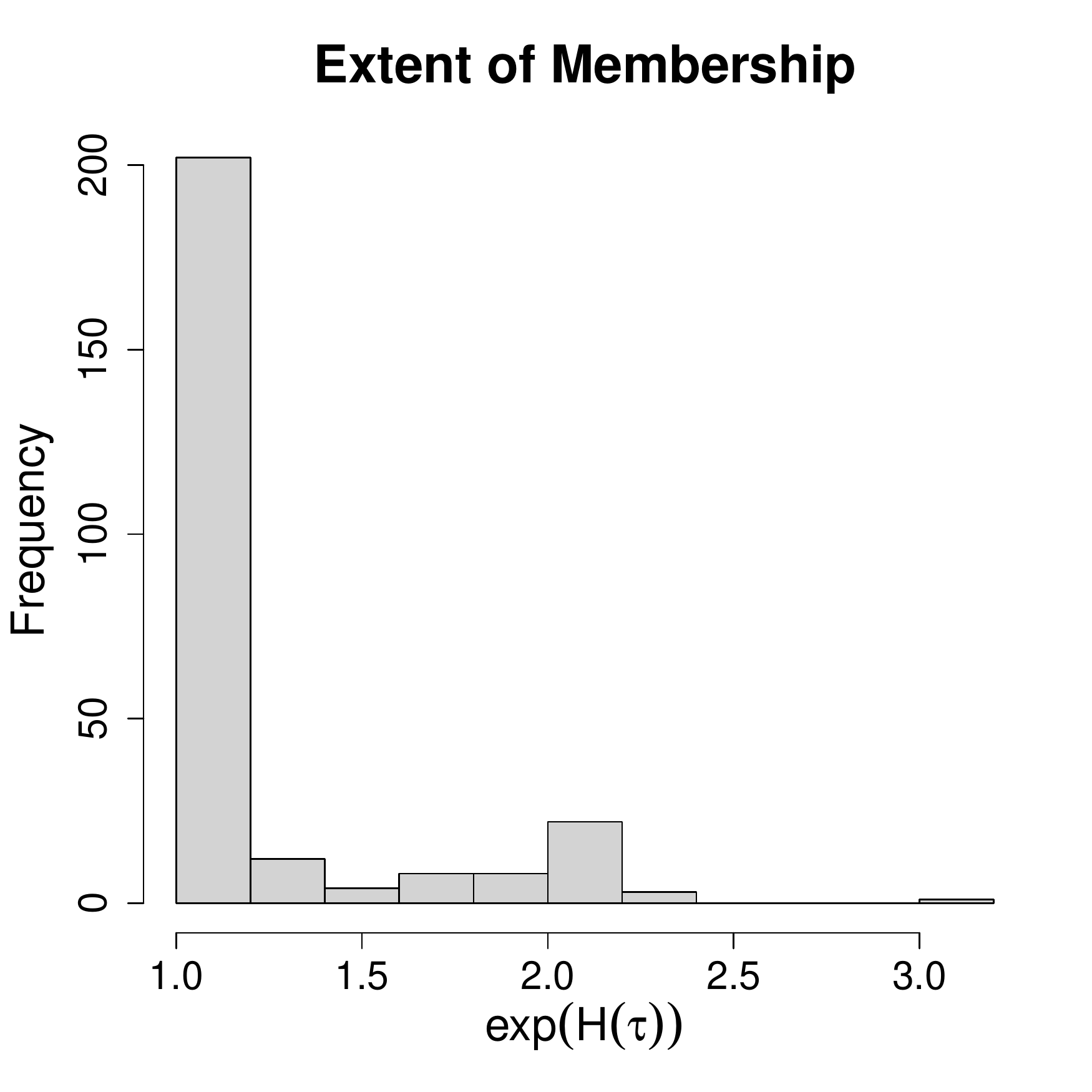}
                \caption{~}
                \label{fig:24H_EoMa}
 \end{subfigure}
\begin{subfigure}[b]{0.49\textwidth}
 \includegraphics[width=0.99\textwidth]{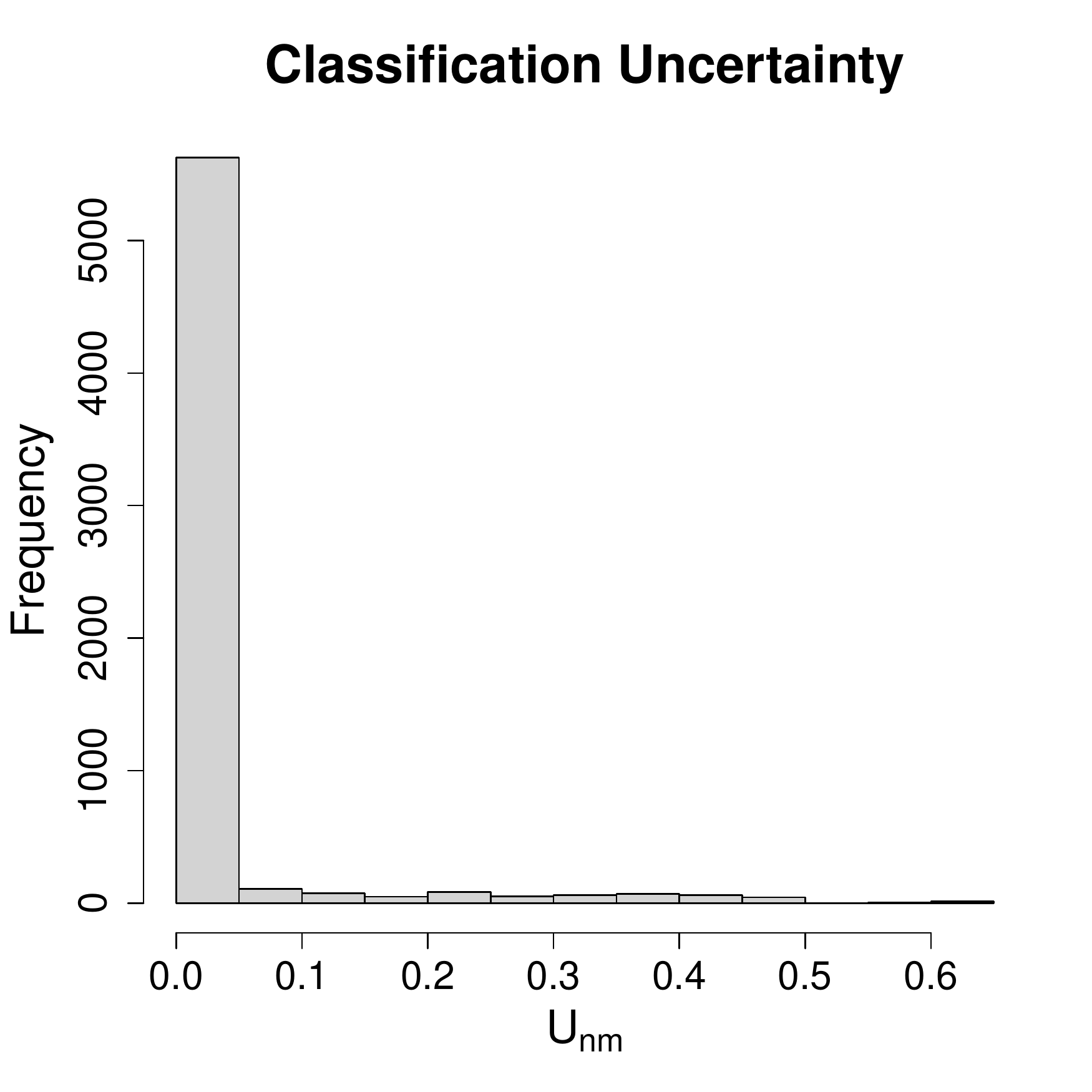}
                \caption{~}
                \label{fig:24H_EoMb}
 \end{subfigure}
\end{center}
\caption{Histograms of the extent of profile membership and classification uncertainty of observations in the running data. Roughly 17\% of observations exhibit membership between 2 profiles, with one observation exhibiting membership to 3 profiles. Over 90\% of datapoints are mapped to a profile with high certainty.}
\label{fig:24H_EoM}
\end{figure}

Figure \ref{fig:24H_EoMa} shows that while the majority of observations exhibit membership to only one profile, about 17\% of observations exhibit at least some mixed membership, with all but one of these observations displaying membership between two profiles. In the mixed membership setting, about 90\% of datapoints are classified with uncertainty less than 5\%, substantially higher than the mixture model clustering, in which only 73\% of observations were clustered with the same level of certainty. Some datapoints are still classified with high uncertainty by the mixed membership clustering; see Figure \ref{fig:24H_EoMb}.

\begin{figure}[th]
\begin{center}
 \begin{subfigure}[b]{0.99\textwidth}
\includegraphics[width=0.99\textwidth]{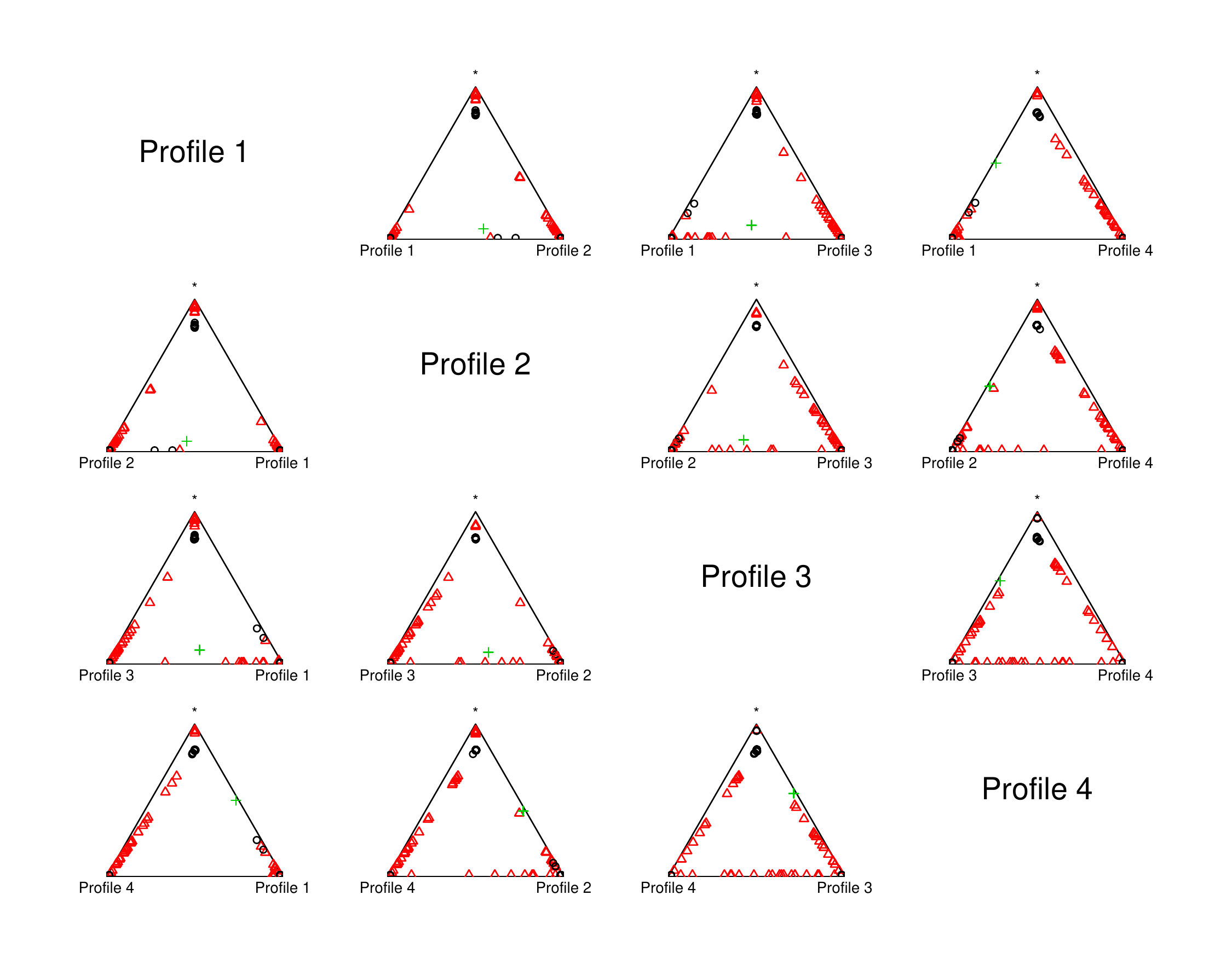}
                \caption{~}
                \label{fig:24H_Tau_Simplexa}
  \end{subfigure}
\end{center}
\caption{Plots of the marginal simplices representing runners' profile membership. Colour and shape are used to denote the number of profiles datapoints map onto. 80\% of observations map to only one profile (black circles), and are closely grouped together. With one exception (the green cross), all remaining datapoints map onto two profiles (red triangles), and are arranged along the edgepoints of the simplices. }
\label{fig:24H_Tau_Simplex}
\end{figure}

\begin{table}[tbh]
\caption{Table detailing which profiles runners map to in mixed membership clustering. Note that the hours of the race which observations map onto different profiles is not provided in this table.  }
\begin{center}
Mapped profile memberships\\
\begin{tabular}{rrrrrrrrrrrr}
  \hline
& \{1\}  & \{2\}  & \{3\}  &  \{4\}  & \{1,2\}  & \{1,3\}   & \{1, 4\}    & \{1, 2, 3\}    & \{2, 3\}   & \{2,4\} & \{3, 4\}  \\ 
  \hline
 & 137 & 42 & 16 & 13  & 1       & 9        & 6         & 1             &  7     &  9        &  19  \\ 
   \hline
\end{tabular}
\end{center}
\label{tab:24HmapZ}
\end{table}

\begin{figure}[th]
\begin{center}
 \begin{subfigure}[b]{0.59\textwidth}
 \includegraphics[width=0.99\textwidth]{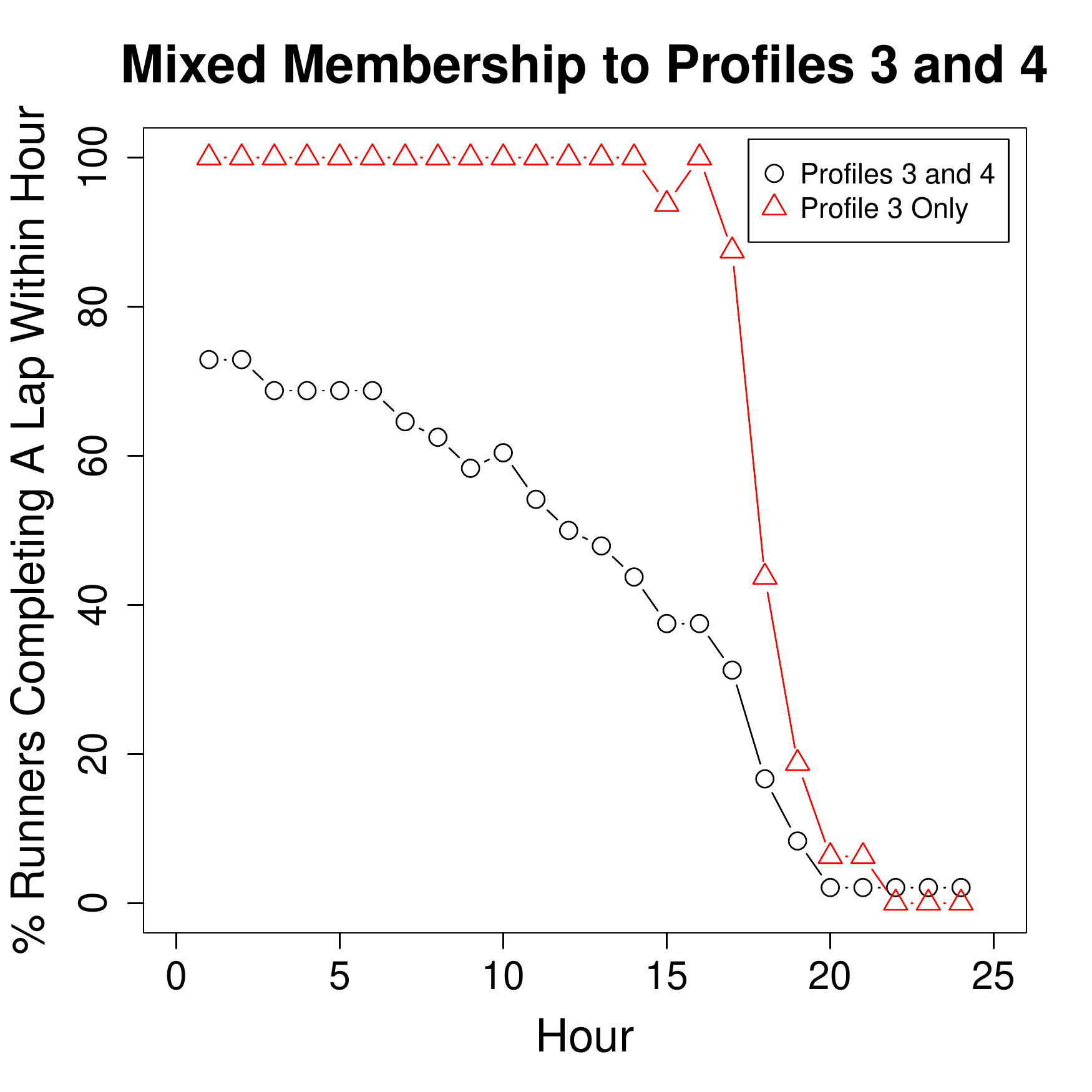}
                \caption{~}
                \label{fig:24H_Tau_Simplexb}
  \end{subfigure}
 \end{center}
 \caption{The percentage of runners in \{3, 4\} and \{3\} who completed at least one lap during each hour of the race.}
  \label{fig:24H_Prof_34}
\end{figure}

A direct inspection of $\mathbf{\hat{Z}}$ shows that a total of 208 of the 260 observations map directly onto one profile, that is, displayed no mixed membership. All except one of the remaining 52 observations display membership across no more than two profiles at one time. The 3-dimensional simplex is visualised using a ternary plot~\citep{vandenBoogaart2008} in Figure~\ref{fig:24H_Tau_Simplexa}. (N.B., recall that we fixed the hyperparameter $\delta_g = 1/4$ in the fitted model.) In cases of mixed membership, the strongest association is between profiles 3 and 4, the Rapid Decline and Non-Participation profiles, as shown in Table \ref{tab:24HmapZ}. The 19 runners exhibiting mixed membership to both these profiles can be characterised as runners starting strongly but whose performance tailed off at various points during the race. While this description is similar to that for the behaviour characterised by the Rapid Decline profile, the behaviour of the two groups is still quite different.  Figure \ref{fig:24H_Tau_Simplexb} shows the percentage of the runners mapped to \{3, 4\}  and the 16 runners mapped to \{3\} who completed at least one lap during each hour of the race; the slope of this line for runners belonging exclusively to the Rapid Decline profile is markedly steeper. This indicates that the pace of runners in \{3, 4\} decline over a much wider time frame.

\subsection{Comparing the Models}

\begin{table}[ht]
\caption{Table comparing clusters found using the mixture model framework to the mapped profiles from the mixed membership. }
\begin{center}
Mapped profile memberships\\
\begin{tabular}{rrrrrrrrrrrr}
  \hline
 & \{1\} & \{1,3\} & \{1,4\} & \{1,2,3\} & \{2\} & \{2,1\} & \{2,3\} & \{2,4\} & \{3\} & \{3,4\} &\{4\} \\ 
  \hline
Group 1 &    98 &   0 &   3 &   0 &  0 &   0 &      0 &   0 &   0 &    0 &   0 \\ 
Group  2 &  39 &   8 &   3 &   1 & 33 &   1 &      4 &   0 &   0 &      0 &   0 \\ 
Group  3 &   0 &   1 &   0 &   0 &  0 &   0 &      3 &   0 &  16 &      2 &   0 \\ 
Group  4 &   0 &   0 &   0 &   0 &  9 &   0 &      0 &   8 &   0 &      0 &   0 \\ 
Group  5 &   0 &   0 &   0 &   0 &  0 &   0 &      0 &   1 &   0 &      2 &  13 \\ 
Group  6 &   0 &   0 &   0 &   0 &  0 &   0 &      0 &   0 &   0 &     15 &   0 \\ 
   \hline
\end{tabular}
\end{center}
\label{tab:24Hcomparison2}
\end{table}

We now compare the clusters found by the mixed membership and mixture modelling frameworks. Table \ref{tab:24Hcomparison2} shows how overlap between the mapped profile memberships from the mixed membership approach compared to the membership of the six groups found using the mixture model framework. Note that 98 of the 101 runners mapped to Group 1 match to \{1\}, the Fast Pace profile. The three runners in the group who exhibit mixed membership do so to \{1,4\}, the Fast Pace and Non-Participation profiles. The runners mapped to these two profiles all ran at a high pace, but failed to complete any laps (possibly stopping completely for that time) for a single hour at different points in the race. Runners clustered together in Group 2 by the mixture model approach are mainly split between \{1\} and \{2\}, the Fast and Slow Pace profiles in the mixed membership approach. The runners in this group exhibiting mixed membership are similar to those with membership of two profiles in Group 1 in that they run at a high pace but stop, or fail to complete a lap, intermittently, before returning to the previous pace. Group 3 corresponds closely to \{3\}, the Rapid Decline profile, while Group 4 matches to  either \{2\} or \{2,4\}, the Slow Pace and Non-Participation profiles, again indicating that some runners in this group raced only intermittently. Members of Group 5 are mainly clustered to \{4\}, the Non-Participation profile, which is perhaps unsurprising. Members of Group 6 are all members of \{3,4\}, the Rapid Decline and Non-Participation profiles;  this behaviour has been discussed in the previous subsection. This indicates that perhaps Group 6, the smallest group in the fitted mixture model, was a poor fit to the data; rather than being a group of runners whose pace gradually decreased, it consisted of a group of runners completing a large number of laps an hour, with various members of the group withdrawing early at different stages in the race. 

\subsection{Examples of Mixed Membership}\label{sec:examples}

\begin{figure*}
    \begin{center}
        \begin{subfigure}[b]{0.4\textwidth}
                \centering
                \includegraphics[width=1.09\linewidth]{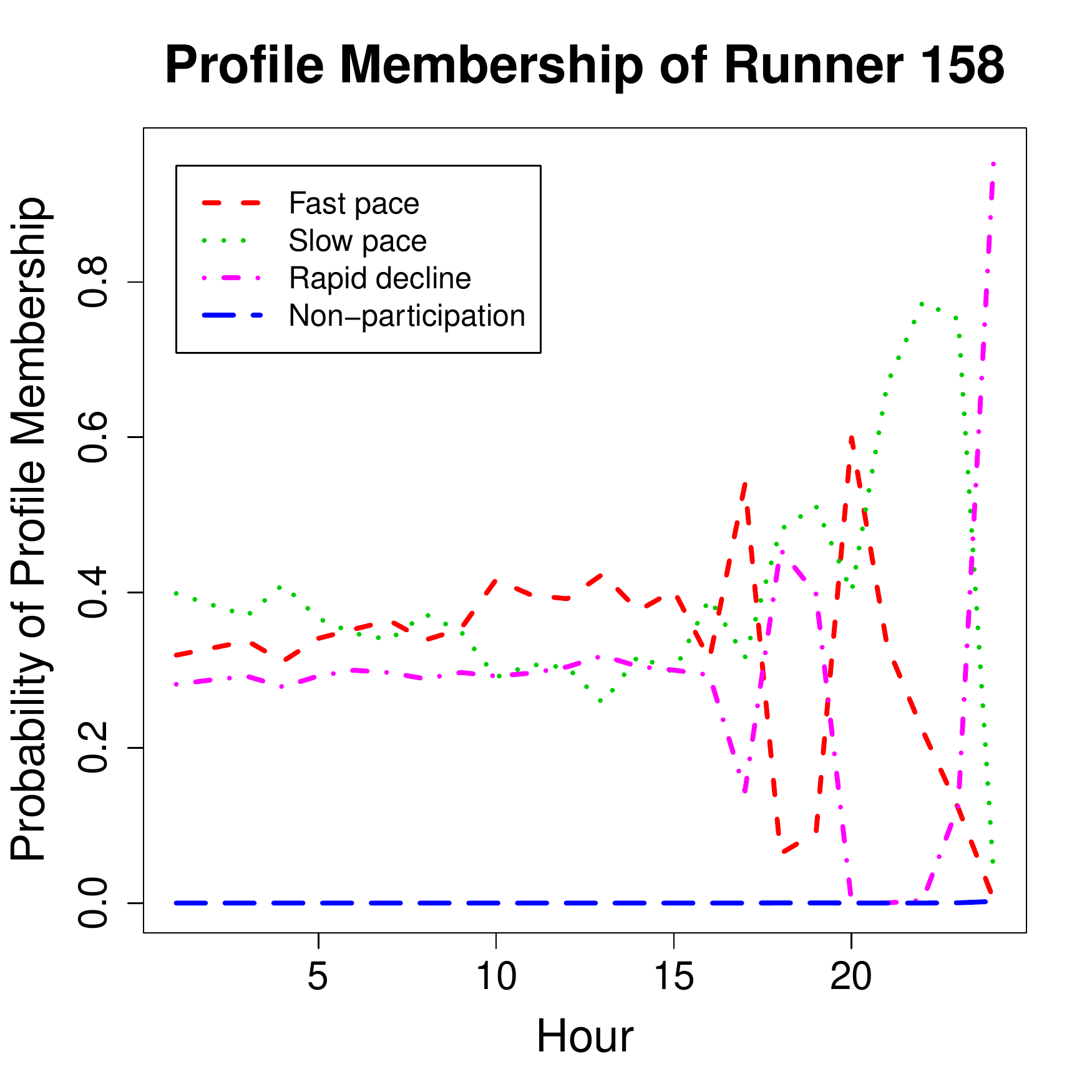}
                \caption{~}
                \label{fig:24HMixMemExamples1a}
        \end{subfigure}
         \begin{subfigure}[b]{0.4\textwidth}
                \centering
                \includegraphics[width=1.09\linewidth]{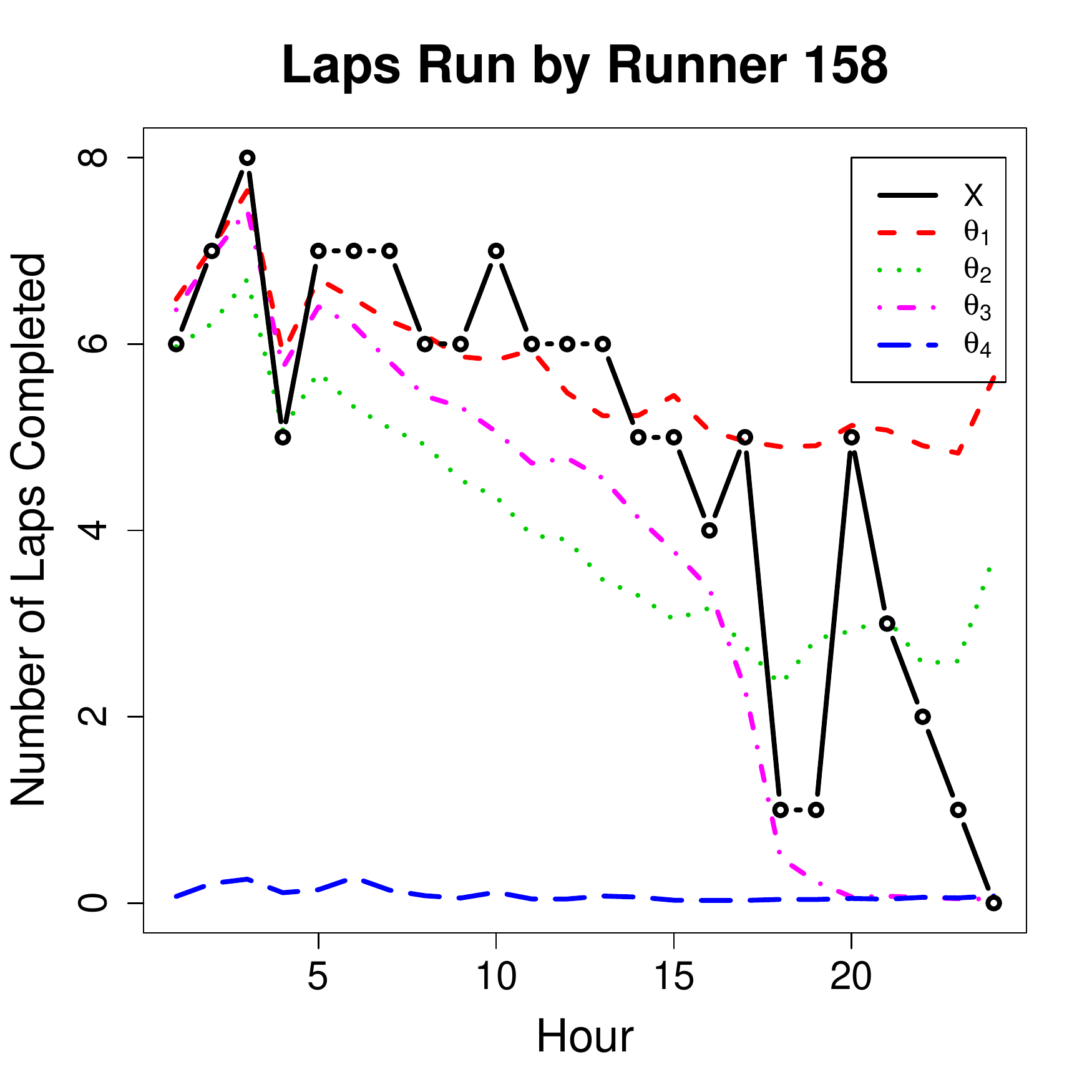}
                \caption{~}
                \label{fig:24HMixMemExamples1b}
        \end{subfigure}
         \vskip 0.8em     
               \begin{subfigure}[b]{0.4\textwidth}
                \centering
                \includegraphics[width=1.09\linewidth]{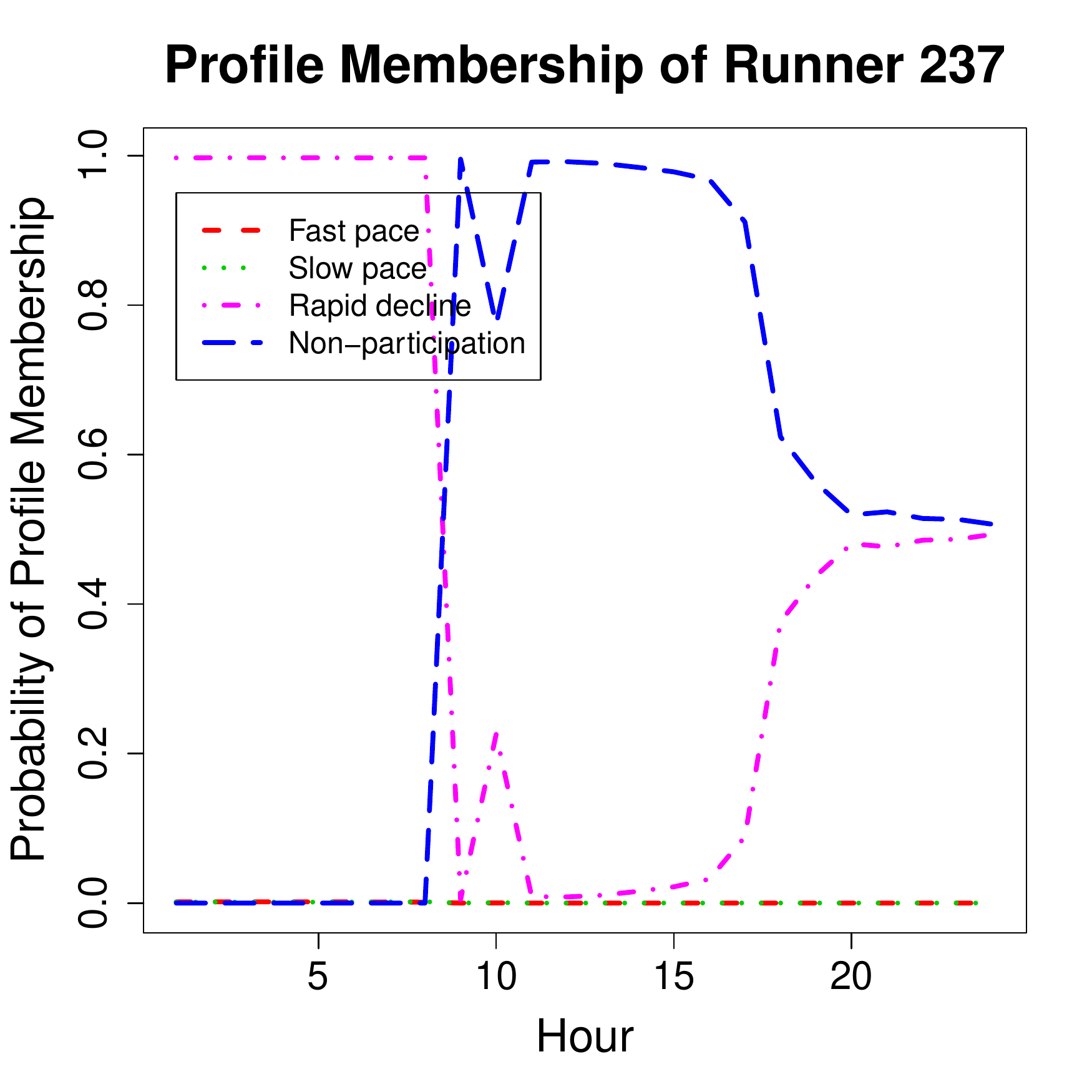}
                \caption{~}
                \label{fig:24HMixMemExamples2a}      
        \end{subfigure}
                \begin{subfigure}[b]{0.4\textwidth}
                \centering
                \includegraphics[width=1.09\linewidth]{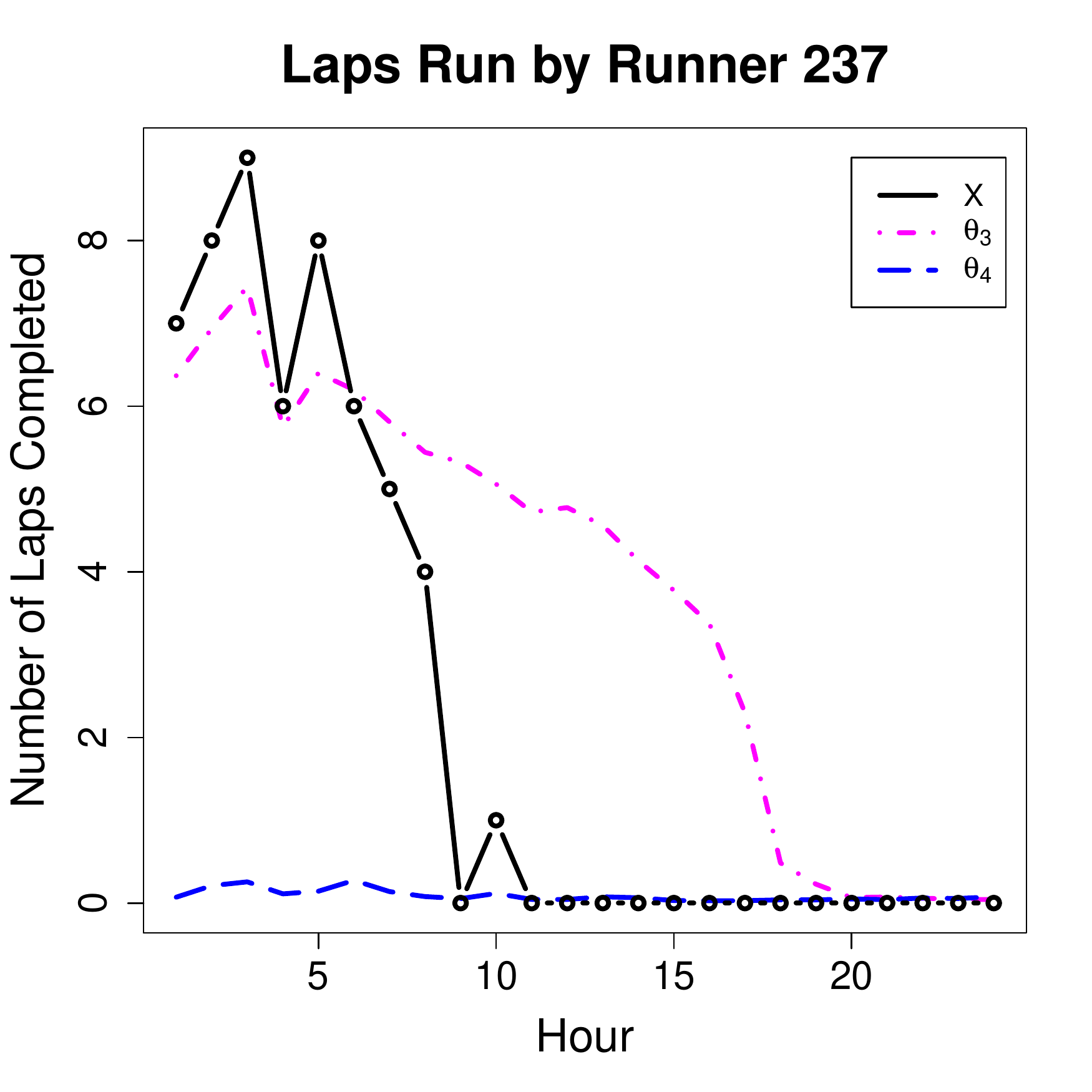}
                \caption{~}
                \label{fig:24HMixMemExamples2b}
        \end{subfigure}
        \vskip 0.8em   
                \begin{subfigure}[b]{0.4\textwidth}
                \includegraphics[width=1.09\linewidth]{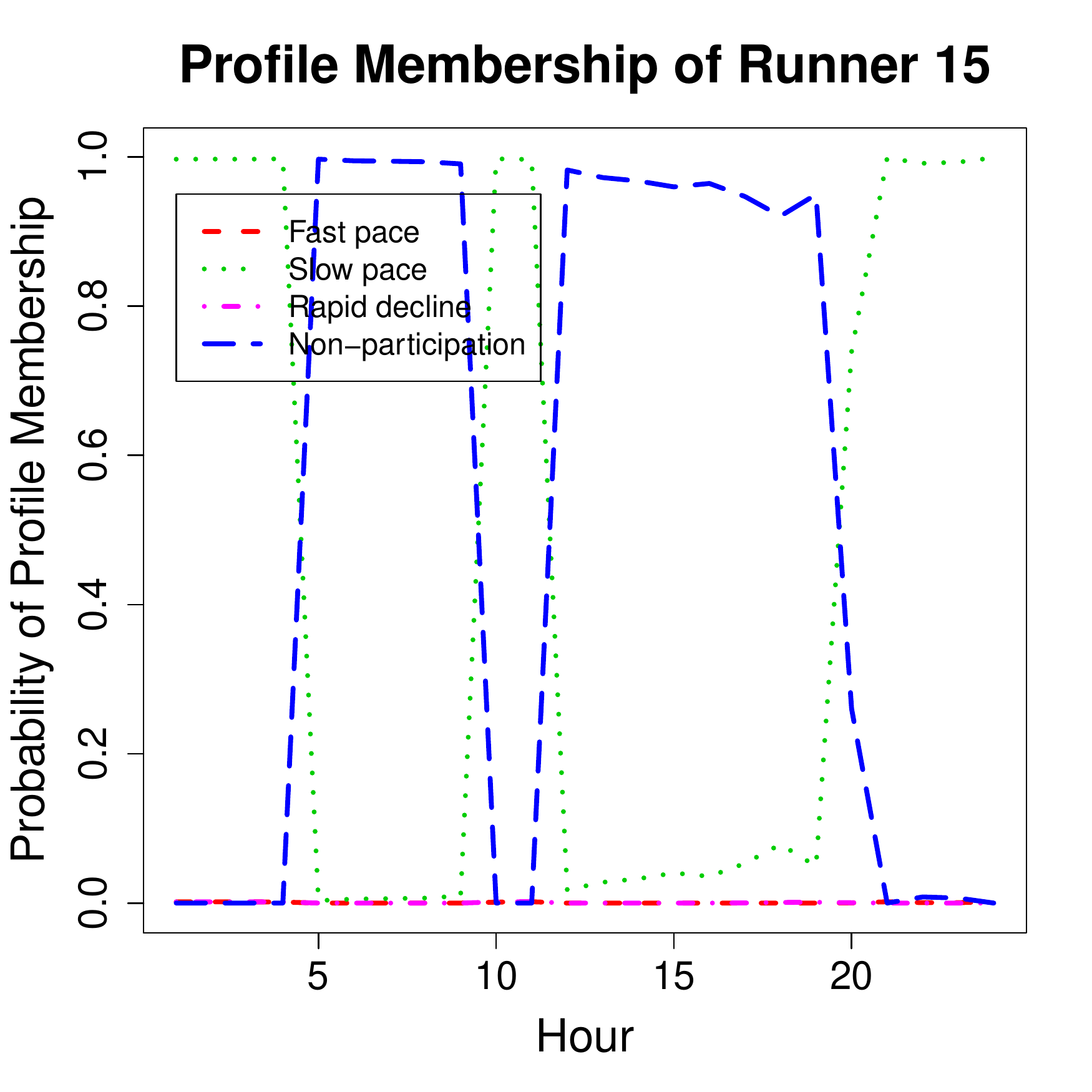}
                \caption{~}
                \label{fig:24HMixMemExamples3a}
        \end{subfigure}
                \begin{subfigure}[b]{0.4\textwidth}
                \centering
                \includegraphics[width=1.09\linewidth]{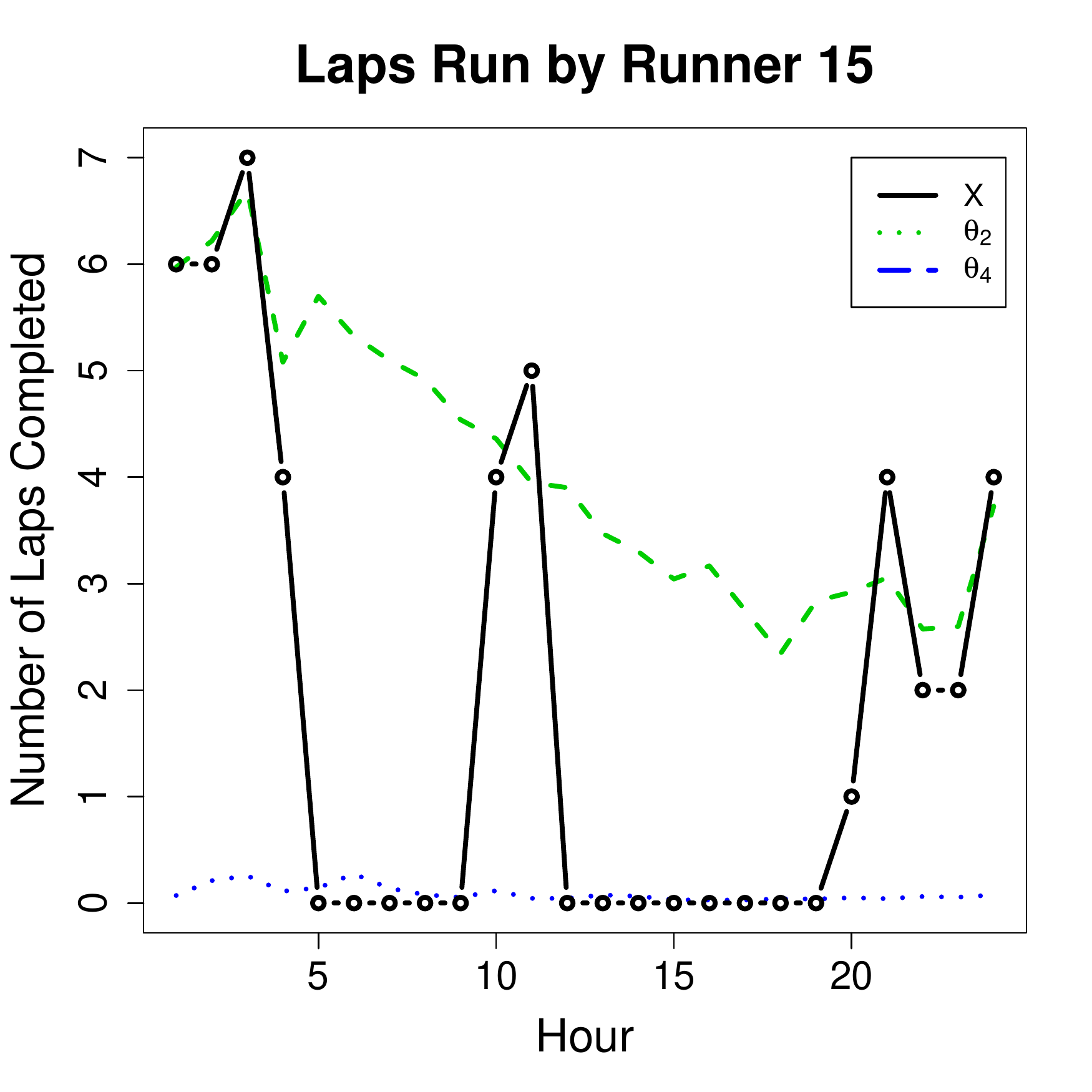}
                \caption{~}
                \label{fig:24HMixMemExamples3b}
        \end{subfigure}
    \end{center}
    \vskip -0.3em
    \caption{Examples of runners exhibiting mixed membership for the 4-profile fit. }
    \label{fig:24HMixMemExamples}
\end{figure*}

In this section, in order to to illustrate the types of mixed membership exhibited by the data, the three race entrants with the highest EoM scores are discussed, in decreasing order. Plots of each runner's lap numbers and profile assignment scores over the course of the race are given in Figure \ref{fig:24HMixMemExamples}.
\begin{description}
\item[\textbf{ Runner 158}] This was the only race entrant to be mapped to three profiles over the course of the race. Inspecting Figure \ref{fig:24HMixMemExamples1a}, it's clear that a high level of uncertainty is associated with this runner's profile membership throughout the race, until the last hour, when their lap time is associated with the Rapid Decline profile with a high level of certainty. Figure \ref{fig:24HMixMemExamples1a} shows the runners data, along with the estimated values  of  $\boldsymbol{\theta}$. From this we can see that for the first half of the race, the runner ran at a good pace, consistent with both the Fast Pace and Rapid Decline profiles. On the 18th hour, this runner experienced a large dip in pace consistent with the Rapid Decline profile, but recovered at hours 21 and 22, again running at a pace more consistent with the Fast and Slow Pace profiles, before eventually fading again for the last two hours.
\item[\textbf{Runner 237}] This runner's performance is characterised as being split between the Rapid Decline and Non-Participation profiles, a type of mixed membership discussed previously. This runner starts well, but does not complete any laps past the 10th hour. Note the high level of uncertainty of profile membership for the last six hours (Figure \ref{fig:24HMixMemExamples2a}); this is explained by the fact that the values of $\boldsymbol{\theta}$ are very close together for Profiles 3 and 4 for these hours, and that this runner has evenly split profile membership between the two profiles for the hours before that in the race.
\item[\textbf{Runner 15}] This runner's profile membership was split between the Slow Pace and Non-Participation profiles. This runner's race can be characterised as running at a at a relatively low pace, while stopping for several hours on two occasions before completing a reasonably high number of laps during the final four hours of the race. Despite this runner's erratic behaviour, given an hour $m$ and profile membership $g$, the number of laps they complete is usually quite close to the value ${\theta}_{gm}$. In this case, the exchangeability assumption of the mixed membership model is arguably advantageous; a model that incorporated too much dependence between race laps could be over smooth by comparison.
\end{description}

\section{Discussion}\label{sec:Conclusions}
It is clear that mixed membership methods provide the analyst with tools of greater flexibility than current MBC or standard distance-based clustering methods. While the mixed membership framework is more elaborate than that of the mixture model, our application makes clear the benefits that the method provides, and that its output can be interpreted and understood. While the nature of the running data seems to be better modelled by a mixed membership approach, at least in a qualitative sense, it is difficult to show this quantitatively, and the question of how to compare different types of clustering method in general remains open.

While in theory it is possible to obtain equivalent clusterings of observations using mixed membership and mixture models, we argue that this is unlikely to occur in practice. For example, in the application to the running data, since several observations have unique profile mappings  -- for example, Runner 15 stops several times --  this would suggest an equivalent clustering solution in the mixture model framework would contain many singleton clusters. Typically such clusterings are considered unfavourable. However within the mixed membership framework, the unique aspects of the runner's behaviour are well explained in this case.


In this paper we have provided a mixed membership formulation for data produced by members of an exponential family with an underlying latent mixed membership structure. It may be of interest to expand this model further to account for mixed-type data, similar to the procedure for mixture models introduced by \citet{Vermunt02}. 
The simplifying assumption of exchangeability made by the model, as discussed in Section \ref{sec:MixMem}, may be somewhat unrealistic; for example, in the running data, runners with partial membership to profile 4 tend to be assigned membership later rather than earlier in the race. While in a general sense, as noted by \citet{blei03}, it may be difficult to justify the epistemological validity of such an assumption, its utility in a clustering framework is clear. In particular, when applied to the running data, the mixed membership approach effectively captures the sporadic nature with which runners stopped throughout the race. 

A potential weakness of the model as currently formulated is the use of the Dirichlet distribution to model each observation's profile membership. The use of this distribution reflects the assumption that the profile membership of an observation's attributes can be thought of as exchangeable entities, causing any correlation within the data to be ignored. Thus the model may have poor posterior predictive power. While not an explicit aim of this paper, it is a limitation of the current model. One solution is to replace the Dirichlet distribution with a logistic normal distribution  \citep{blei07} although this complicates the inference method. \citet{Wang2013} have outlined methods for performing inference in a variational Bayes setting when the posterior form is non-conjugate. Additionally, longitudinal mixed membership models have been developed. \citet{ManriqueVallier2014} explicitly models profile behaviour as a function of time, while \citet{blei2006} allow profile behaviour and the \emph{a priori} probability of profile membership to evolve over time using a state space approach. 


\begin{acknowledgement}
This work is supported by Science Foundation Ireland under the Clique Strategic Research Cluster (08/SRC/I1407) and Insight Research Centre grant (SF1/12/RC/2289).
\end{acknowledgement}

\bibliography{writeup}
\end{document}